\documentstyle[prd,aps,floats]{revtex}

\def\R{{\cal R}}

\def\be{\begin{equation}}
\def\ee{\end{equation}}
\def\bea{\begin{eqnarray}}
\def\eea{\end{eqnarray}}
\def\Aa{\frac{a^{\prime}}{a}}
\def\Ab{\Big(\frac{a^{\prime}}{a}\Big)^2}
\def\Ac{\frac{a^{\prime\prime}}{a}}

\def\La{\partial_i \,\partial^i}
\def\LA{\partial_k \,\partial^k}
\def\deu{{\delta}^{(1)}}
\def\ded{{\delta}^{(2)}}

\begin{document}
\draft

%
%
%
\renewcommand{\topfraction}{0.99}
\renewcommand{\bottomfraction}{0.99}
\title{Second-Order Cosmological Perturbations from Inflation}
\author{Viviana Acquaviva$^1$\footnote{Address after Nov. 2002: SISSA/ISAS, Via
Beirut 4, I-34014, Trieste, Italy.}, Nicola Bartolo$^{2,3}$
\footnote{Address after Nov. 2002: Astronomy Centre, University of Sussex,
Brighton BN1 9QJ, U.K..},
Sabino Matarrese$^{2,3}$ and Antonio Riotto$^3$}
\address{(1)Dipartimento di Fisica, Universit\`a di Pisa, via Buonarroti 2, 
I-56100, Pisa, Italy}
\address{(2)  Dipartimento di Fisica di Padova ``G. Galilei'',
Via Marzolo 8, Padova I-35131, Italy}
\address{(3)  INFN, Sezione di Padova, Via Marzolo 8, Padova I-35131, Italy}

\date{\today}
\maketitle
\begin{abstract}
We present  the first computation of the cosmological perturbations
generated during inflation up to second order in deviations from the
homogeneous background solution. Our results, which fully account for the
inflaton self-interactions as well as for the second-order fluctuations
of the background metric, provide the exact
expression for the gauge-invariant curvature perturbation bispectrum
produced during inflation in terms of the slow-roll parameters or,
alternatively, in terms of the scalar spectral  $n_S$ and
and the tensor to adiabatic scalar amplitude ratio $r$.
The bispectrum represents a specific non-Gaussian signature of
fluctuations
generated by quantum oscillations during slow-roll inflation. However,
our findings indicate that detecting 
the  non-Gaussianity in the cosmic 
microwave background anisotropies emerging from
the second-order calculation will be a challenge for   the
forthcoming satellite experiments.

\end{abstract}

\pacs{PACS numbers: 98.80.Cq \hfill DFPD--A-02-21,
astro-ph/0209156}

\vskip2pc
\section{Introduction}
Inflation represents a successful  mechanism for the causal generation
of primordial  cosmological perturbations
in the early Universe \cite{lr}.
These fluctuations  are then amplified by the gravitational
instability to seed structure formation in the Universe, and to produce
Cosmic Microwave Background (CMB) anisotropies.
Since the primordial cosmological perturbations are 
tiny, the generation and evolution of fluctuations during inflation has always
been studied within linear theory. 
On the other hand, there exist physical observables, such as the
three-point function of scalar perturbations, or its Fourier transform, the
bispectrum, for which a perturbative treatment up to second order
is required, in order to obtain a self-consistent result

The importance of the bispectrum comes from the fact that it represents
the lowest order statistics able to distinguish non-Gaussian from Gaussian
perturbations for which odd-order correlation functions necessarily
vanish.
An accurate calculation of the primordial bispectrum of cosmological
perturbations has become an extremely important issue, as a number of
present
and future experiments, such as MAP and {\it Planck}, will allow to
constrain
or detect non-Gaussianity of CMB anisotropy data with high precision.

So far, the problem of calculating the bispectrum of perturbations
produced
during inflation has been addressed by either looking at the effect of
inflaton self-interactions (which necessarily generate non-linearities in
its
quantum fluctuations) in a fixed de Sitter background \cite{FRS}, or by
using
the so-called stochastic approach to inflation \cite{Getal}\footnote{
See also Ref. \cite{wk}.}, where
back-reaction effects of field fluctuations on the background metric are
partially taken into account. An intriguing result of the stochastic
approach
is that the dominant source of non-Gaussianity actually comes from
non-linear
gravitational perturbations, rather than by inflaton self-interactions.

In this paper we provide for the first time the computation of the scalar
perturbations produced during single-field slow-roll inflation up to
second
order in deviations from the homogeneous background. We achieve different
goals. First, we  provide a gauge-invariant definition of the comoving
curvature ${\cal R}$ at second order of perturbation theory.
The importance of the  second-order comoving
curvature perturbation ${\cal R}$ comes from the fact that it allows
to compute the 
three-point correlation function for the primordial
scalar perturbations --  or its Fourier transform, the 
bispectrum -- which represent 
the lowest order statistics able to distinguish non-Gaussian from Gaussian 
perturbations.
Secondly, we
show that the second-order comoving curvature perturbation is conserved on
super-horizon scales, like its first-order counterpart\footnote{
For related results see  also Ref. \cite{sal}.}. Third, we obtain
the
expression for the gauge-invariant gravitational potential
bispectrum
during inflation, in terms of slow-roll parameters or, equivalently, of
the
spectral indices of the scalar and tensor power-spectra.

An accurate calculation of the primordial bispectrum of cosmological 
perturbations has become a crucial  issue, as a number of present 
and future experiments, such as MAP and {\it Planck}, will allow to constrain 
or detect non-Gaussianity of CMB anisotropy data with high precision.

The plan of the paper is as follows. In Section II 
we write the perturbations of the metric for 
a spatially flat Robertson-Walker background up to second order 
and we derive consistently the fluctuations  
of the energy-momentum tensor of a scalar field. In Section III 
we demonstrate how to find 
a second-order gauge-invariant definition of the comoving 
curvature perturbation. Section IV will be devoted 
to the perturbed Einstein equations up to second order 
in the metric and in the inflaton fluctuations. 
We shall explain how to derive the evolution of the curvature 
perturbation on large scales during a period 
of cosmological inflation by performing an expansion to 
lowest order in the slow-roll parameters. 
 Finally, in Section V 
we draw some concluding remarks relating our findings to the 
gauge-invariant gravitational potential 
bispectrum which is the main physical observable which carries information
about the primordial non-Gaussianity. We also provide two Appendices 
where the reader can find various  technical details.

\section{Perturbations of a flat Robertson-Walker Universe up to second order}
We first write down the perturbations on a spatially flat 
Robertson-Walker background following the formalism of Refs.\cite{BMMS,MMB}. 
We shall first consider the fluctuations of the metric, and then the 
fluctuations of the energy-momentum tensor of a scalar field. Hereafter 
greek indices run from $0$ to $3$, while latin indices label  
the spatial coordinates from $1$ to $3$. If not otherwise specified we 
will work with conformal time $\tau$, and a prime will stand for a derivative
with respect to $\tau$.
\subsection{The metric tensor}
The components of a perturbed spatially flat Robertson-Walker 
metric can be written as
\bea \label{metric1}
g_{00}&=&-a^2(\tau)\left( 1+2 \phi^{(1)}+\phi^{(2)} \right)\, ,\nonumber\\
g_{0i}&=&a^2(\tau)\left( \hat{\omega}_i^{(1)}+\frac{1}{2} 
\hat{\omega}_i^{(2)} \right)
\, ,  \nonumber\\g_{ij}&=&a^2(\tau)\left[
(1 -2 \psi^{(1)} - \psi^{(2)})\delta_{ij}+
\left( \hat{\chi}^{(1)}_{ij}+\frac{1}{2}\hat{\chi}^{(2)}_{ij} \right)\right] 
\,.
\eea 
The standard splitting of the perturbations into scalar, transverse 
({\it i.e} divergence-free) vector parts, and transverse trace-free tensor 
parts with respect to the 3-dimensional space with metric $\delta_{ij}$, 
can be performed in the following way:
\be
\hat{\omega}_i^{(r)}=\partial_i\omega^{(r)}+\omega_i^{(r)}\, ,
\ee
\be
\hat{\chi}^{(r)}_{ij}=D_{ij}\chi^{(r)}+\partial_i\chi^{(r)}_j+\partial_j\chi^{(r)}_i
+\chi^{(r)}_{ij}\, ,
\ee
where $(r)=(1),(2)$ stand for the order of the perturbations, $\omega_i$
and $\chi_i$ are transverse vectors ($\partial^i\omega^{(r)}_i=
\partial^i\chi^{(r)}_i=0$), $\chi^{(r)}_{ij}$ is a symmetric transverse and 
trace-free tensor ($\partial^i\chi^{(r)}_{ij}=0$, $\chi^{i(r)}_{~i}
=0$) and $D_{ij}=\partial_i \partial_j - (1/3) \, \, 
\delta_{ij}\, \partial^k\partial_k$ is a trace-free \break operator 
\footnote{Notice that our notation is different from that of Refs. 
\cite{mfb,malik} for the presence of $D_{ij}$, 
while it is closer to the one used in Refs.~\cite{bardeen,KS}. 
As far as the first-order perturbations are concerned, the metric 
perturbations 
$\psi$ and $E$ of Refs.~\cite{mfb,malik} are 
given in our notation as $\psi=\psi^{(1)}+(1/6)\, \La \chi^{(1)}$ and 
$E=\chi^{(1)}/2$, respectively. However, no difference appears 
in the calculations when using the generalized longitudinal gauge in 
Eq. (\ref{metric3}).}. Here and in the following latin indices 
are raised and lowered using $\delta^{ij}$ and $\delta_{ij}$, respectively.\\
For our purposes the metric in Eq.~(\ref{metric1}) can be simplified. In fact, 
first-order vector perturbations are zero in the presence of a scalar field; 
moreover, the tensor part gives a negligible contribution to the bispectrum.
Thus, in the following we can neglect 
$\omega^{(1)}_i$, $\chi^{(1)}_i$ and $\chi^{(1)}_{ij}$.
However the same is not true for the second order perturbations. 
In the second-order theory the second-order vector and tensor 
contributions can be generated by the first-order scalar perturbations 
even if they are initially zero 
\cite{MMB}. Thus we have to take them into account and we shall use the metric
\bea \label{metric2}
g_{00}&=&-a^2(\tau)\left( 1+2 \phi^{(1)}+\phi^{(2)} \right)\, ,\nonumber\\
g_{0i}&=&a^2(\tau)\left( \partial_i\omega^{(1)}+\frac{1}{2}\, 
\partial_i\omega^{(2)}+\frac{1}{2}\, \omega_i^{(2)} \right)
\, ,  \nonumber\\g_{ij}&=&a^2(\tau)\left[
\left( 1 -2 \psi^{(1)} - \psi^{(2)} \right)\delta_{ij}+
D_{ij}\left( \chi^{(1)} +\frac{1}{2} \chi^{(2)} \right)
+\frac{1}{2}\left( \partial_i\chi^{(2)}_j
+\partial_j\chi^{(2)}_i
+\chi^{(2)}_{ij}\right)\right]
\,.
\eea 
The controvariant metric tensor is obtained by requiring (up to second order)
that $g_{\mu\nu}g^{\nu\lambda}=\delta_\mu\, ^\lambda$ and it is given by
\bea \label{cont}
g^{00}&=&-a^{-2}(\tau)\left( 1-2 \phi^{(1)}-\phi^{(2)} +4\left(
\phi^{(1)}\right)^2-\partial^i\omega^{(1)}\partial_i\omega^{(1)}  
\right)\, ,\nonumber\\
g^{0i}&=&a^{-2}(\tau)\left[ \partial^i\omega^{(1)}+\frac{1}{2}
\left( \partial^i\omega^{(2)}+\omega^{i(2)} \right) +2 \left( \psi^{(1)}
-\phi^{(1)} \right) \partial^i\omega^{(1)}-\partial^i\omega^{(1)}
D^i\,_k \chi^{(1)} \right]
\, ,  \nonumber\\
g^{ij}&=&a^{-2}(\tau) \left[
\left( 1+2 \psi^{(1)} +\psi^{(2)}+4 \left( \psi^{(1)} \right)^2 
\right) \delta^{ij}-
D^{ij}\left( \chi^{(1)} +\frac{1}{2} \chi^{(2)} \right)\right.\nonumber \\
&-&\left. \frac{1}{2}\left( \partial^i\chi^{j(2)}
+\partial^j\chi^{i(2)}
+\chi^{ij(2)} \right)-\partial^i\omega^{(1)}\partial^j\omega^{(1)}
-4 \psi^{(1)}D^{ij}\chi^{(1)}+D^{ik}\chi^{(1)}D^j_{~k} \chi^{(1)} \right]\,.
\eea
Using $g_{\mu\nu}$ and $g^{\mu\nu}$ one can calculate the 
connection coefficients and the Einstein tensor components up to second order in the metric 
fluctuations. 
Their complete expressions are contained in Appendix \ref{A}.  

\subsection{Energy-momentum tensor of a scalar field}
We shall consider a scalar field $\varphi(\tau, x^{i})$ 
minimally coupled to gravity, whose energy-momentum tensor is given by
\be \label{EM}
T_{\mu\nu}\,=\,\partial_\mu \varphi\,\partial_\nu \varphi \,-\,
g_{\mu\nu}\left( \frac{1}{2}\,g^{\alpha\beta}\,\partial_\alpha
\varphi\,\partial_\beta \varphi \,+\, V(\varphi)\right)\, ,
\ee    
where $V(\varphi)$ is the potential of the scalar field. A successfull 
period of inflation can be attained when the potential  $V(\varphi)$
 is flat enough. 

The scalar field can be split into a homogeneous background $\varphi_0(\tau)$ 
and a perturbation $\delta \varphi (\tau, x^{i})$ as
\be \label{scalarfield}
\varphi(\tau, x^i)=\varphi_0(\tau)+\delta\varphi(\tau, x^i)=\varphi_0(\tau)+
\delta^{(1)}\varphi(\tau, x^i)+\frac{1}{2}\delta^{(2)}
\varphi(\tau, x^i)\, ,
\ee
where the perturbation has been expanded into a first and a second-order
part, respectively. Using the expression (\ref{scalarfield}) into Eq. (\ref{EM}) and calculating
$T^\mu_{~\nu}=g^{\mu\alpha}T_{\alpha\nu}$ up to second order we find 
\bea 
T^\mu_{~\nu}=T^{\mu(0)}_{~\nu}+\deu T^{\mu}_{~\nu}+\frac{1}{2}\ded
T^{\mu}_{~\nu} \, ,
\eea
where $T^{\mu(0)}_{~\nu}$ corresponds to the background value, and
\bea 
\label{scalT_00}
T^{0(0)}_{~0}+\deu T^{0}_{~0}&=& -{1\over2}a^{-2} \varphi'^{~2}_0 - V_0  
+ a^{-2} \varphi_0' \left(\phi^{(1)}~\varphi_0'-  {\deu{\varphi}}^{\prime} \right) 
-\frac{\partial V}{\partial \varphi} \deu\varphi \, , \\ 
\label{00}
\ded T^0_{~0}&=& \frac{2}{a^2} \left[-\frac{1}{2}{\ded{\varphi}}^{\prime}\,\varphi_0^{\prime}  \,-\,
\frac{1}{2}\ded{\varphi} \,\frac{\partial V}{\partial \varphi} \,a^2 \,+\,
\frac{1}{2}\,\phi^{(2)}\,{\varphi_0^{\prime}}^2\,-\,
\frac{1}{2}\,\left( {\deu{\varphi}}^{\prime} \right)^2 \,-\, \frac{1}{2}
\partial^k \deu{\varphi} \,\partial_k \,\deu{\varphi} \right.\\
&-&\left. \frac{1}{2}\,\left( \deu{\varphi} \right )^2 \,\frac{\partial^2 V}{\partial
\varphi^2} \,a^2 \,-\, 2\,\left( \phi^{(1)} \right)^2\,{\varphi_0^{\prime}}^2 \,+\,
2\,\phi^{(1)}\,{\deu{\varphi}}^{\prime}\,\varphi_0^{\prime} \,+\,
\frac{1}{2}\,\partial^k \omega^{(1)} \,\partial_k \omega^{(1)} \,{\varphi_0^{\prime}}^2\right]
\, , \nonumber \\
&& \nonumber \\
&& \nonumber \\
\label{scalT_0i}
T^{0(0)}_{~i}+\deu T^{0}_{~i}&=& -a^{-2} \left(\varphi_0^{\prime} \,  
\partial_i\deu\varphi \right) \, , \\
\label{0i}
\ded {T^0_{~i}}&=&\frac{2}{a^2} \left(-\frac{1}{2}\varphi_0^{\prime}\, \partial_i
\ded{\varphi}-\,
\partial_i \,\deu{\varphi} \, {\deu{\varphi}}^{\prime} \,+\, 2\,\varphi_0^{\prime}\,
\phi^{(1)}\, 
\partial_i \deu{\varphi} \right)\, ,\\
&& \nonumber \\
&& \nonumber \\
T^{i(0)}_{~0}+\deu T^{i}_{~0}&=& {\varphi_0^{\prime}}^2 \partial^{i}\omega^{(1)}+\varphi_0^{\prime}\,
\partial^{i}\deu\varphi \, , \\
\label{i0}
\ded {T^i_{~0}}&=& \frac{2}{a^2} \left[ \frac{1}{2}\varphi_0^{\prime}\,\partial^i \,\ded{\varphi} \,+\,
\frac{1}{2}\,\left( \partial^i \omega^{(2)} + \omega^{i(2)} \right) {\varphi_0^{\prime}}^2 \,+\,
{\deu{\varphi}}^{\prime} \partial^i \,\deu{\varphi}\,+\,
2\,\varphi_0^{\prime}\, \psi^{(1)} \,\partial^i \,\deu{\varphi}  \right. \\
&-&\left. 2\, {\varphi_0^{\prime}}^2\, \phi^{(1)}\,
\partial^i \omega^{(1)} \,
+\, 2\,\partial^i \omega^{(1)} \,{\deu{\varphi}}^{\prime}\,\varphi_0^{\prime} \, +\,
2\,{\varphi_0^{\prime}}^2 \, \psi^{(1)}\,\partial^i \omega^{(1)} \,-\, \varphi_0^{\prime}\,D^{ij}\chi^{(1)}
\,\partial_j\,\deu{\varphi} \nonumber \right. \\
&-& {\varphi_0^{\prime}}^2\, \left. D^{ij}\chi^{(1)} \,\partial_j \omega^{(1)} \right]\, , \nonumber \\
&& \nonumber \\
&& \nonumber \\
T^{i(0)}_{~j}+\deu T^{i}_{~j}&=&\left[ {1\over2}a^{-2}  {\varphi_0^{\prime}}^2 - V_0 
- \frac{\partial V}{\partial \varphi} \deu\varphi +a^{-2} \varphi_0^{\prime} 
\left(\deu\varphi^{\prime} -\phi^{(1)}~\varphi_0^{\prime} \right) \right]~\delta^i_{~j} \,, \\
\label{ij}
\ded{T^i_{~j}}&=& \frac{2}{a^2} \left[\left(\frac{1}{2} 
{\ded{\varphi}}^{\prime}\,\varphi_0^{\prime} 
\,-\, \frac{1}{2}\ded{\varphi}\,\frac{\partial V }{\partial \varphi}\,a^2 \,-\,
\frac{1}{2}\,\phi^{(2)}\,{\varphi_0^{\prime}}^2 \,+\,
\frac{1}{2}\,\left( {\deu{\varphi}}^{\prime} \right)^2 \,-\,
\frac{1}{2}\,\partial_k
\,\deu{\varphi}\,\partial^k \deu{\varphi} \right. \right. \\
&+& \left.  \left. 2\,\left( \phi^{(1)} \right)^2\,{\varphi_0^{\prime}}^2 \,-\, \frac{1}{2}\, \left(
\deu{\varphi} \right)^2
\,\frac{\partial^2 V}{\partial \varphi^2}\,a^2  \,-\,
2\,\phi^{(1)}\,{\deu{\varphi}}^{\prime} \,\varphi_0^{\prime}\,-\,
\partial^k \omega^{(1)}\,\partial_k \deu{\varphi}\,\varphi_0^{\prime} \right. \right. \nonumber \\
&-& \left. \left. \frac{1}{2}\,\partial^k \omega^{(1)}\,\partial_k \omega^{(1)}\,{\varphi_0^{\prime}}^2
\right)\,\delta^i_{~j}\,+\,\partial^i \deu{\varphi} \,\partial_j
\deu{\varphi} \,+\varphi_0^{\prime}\,
\partial^i
\omega^{(1)}\,\partial_j\,\deu{\varphi} \right]\, , \nonumber 
\eea
with $V_0=V(\varphi_0)$. A comment is in order here. As it can be seen from Eq.~(\ref{cont}) 
and Eqs.~(\ref{00}),(\ref{0i}),(\ref{i0}) and (\ref{ij}) 
the second-order perturbations always contain two different contributions, 
quantities which are intrinsically of second order, 
and quantities which are given by the product of two first-order 
perturbations. As a consequence, when considering the 
Einstein equations to second order in Section \ref{EE}, 
first-order perturbations behave as a source for the 
intrinsically second-order fluctuations. This is an important issue which was pointed out in different 
works on second-order perturbation theory \cite{Tomita,MPS,MMB} 
and it plays a central role in deriving our main results on 
the primordial non-Gaussianity.

\section{The second-order gauge-invariant comoving curvature perturbation}
Let us now focus on the primordial cosmological perturbations produced during 
a period of inflation 
driven by the inflaton $\varphi$. The fluctuations of the inflaton produce an adiabatic density 
perturbation which is associated with a perturbation of the spatial curvature $\psi$. 
The density/curvature 
perturbation can be defined in a gauge-invariant manner as the curvature perturbation $\cal{R}$ 
on slices orthogonal to comoving wordlines. At first order, the comoving curvature perturbation 
is given 
by the gauge-invariant formula \cite{mfb}
\be \label{R1}
{\cal R}^{(1)}= \psi^{(1)}+\frac{{\cal H}}{\varphi_0^\prime}\,
\delta^{(1)}\varphi\, ,
\ee
where ${\cal H}=a^\prime/a$ is the Hubble rate,
$\psi^{(1)}$ is the gauge-dependent first-order curvature perturbation 
and $\deu \varphi$ is the first-order inflaton perturbation in that gauge. An important feature 
of ${\cal R}^{(1)}$ is that it remains constant on super-Hubble scales while approaching the 
horizon entry.\\
We want to find the gauge-invariant comoving curvature perturbation ${\cal R}$ up to second order.
Thus we expand the curvature perturbation $\psi$ and the inflaton perturbation $\delta\varphi$
as $\psi=\psi^{(1)}+\frac{1}{2}\,\psi^{(2)}$ and $\delta\varphi=\delta^{(1)}\varphi+ \frac{1}{2}\, 
\delta^{(2)}\varphi$, according to Eqs. (\ref{metric1}) and (\ref{scalarfield}).
An infinitesimal coordinate change
up to second order induces a gauge transformation of the metric and 
the scalar field perturbations 
\cite{BMMS,MMB}. In particular for a second-order shift of the time coordinate
\be \label{timechange}
\tau\rightarrow \tau-\xi^0_{(1)}+\frac{1}{2} \left({\xi^{0'}_{(1)}}\xi^0_{(1)}
-\xi^0_{(2)}\right)\,
\ee
$\psi^{(2)}$ and $\ded \varphi$ will transform 
as \cite{BMMS,MMB}
\bea \label{T1}
\widetilde{{\psi}^{(2)}} & = &\psi^{(2)}+2 \xi^0_{(1)} \left( {\psi^{(1)}}^{\prime}
+2 {\mathcal{H}} \psi^{(1)} \right) 
-  \left({\mathcal{H}}' +2 {\mathcal{H}}^2
\right) \left(\xi^{0}_{(1)}\right)^2
-  {\mathcal{H}}{\xi_{(1)}^{0}}^{\prime}
\xi_{(1)}^{0} - {\mathcal{H}} \xi^0_{(2)} 
-\frac{1}{3}\left( 2 \partial^i\omega^{(1)} - \partial^i \xi^{0}_{(1)} 
\right)\partial_i \xi^{0}_{(1)}  
\, , \\
\label{T2}
\widetilde{\delta^{(2)}\varphi}& = &\delta^{(2)}\varphi+\xi^0_{(1)}
\left(\varphi''_{0}
\xi^0_{(1)} +\varphi'_{0}{\xi^{0}_{(1)}}^\prime
+2{\delta^{(1)}\varphi}^{\prime}\right) + \varphi'_{0}\xi^0_{(2)} \, .
\eea 
We find that the gauge-invariant comoving curvature perturbation
${\cal R}={\cal R}^{(1)}+\frac{1}{2}{\cal R}^{(2)}$
is provided by 
\bea \label{R}
{\cal R} = {\cal R}^{(1)}+\frac{1}{2} \left[ {\mathcal{H}}\frac{\delta^{(2)}
\varphi}
{{\varphi'_0}}+ \psi^{(2)} \right]
+ \frac{1}{2} \frac{ \left( {\psi^{(1)}}^{\prime}+2{\mathcal{H}}\psi^{(1)}
+{\mathcal{H}} {\deu\varphi}^{\prime}/ \varphi'_0 \right)^2}{{\mathcal{H}}'
+2{\mathcal{H}}^2-{\mathcal{H}}\, \varphi''_0/\varphi'_0}
-\frac{1}{6} \partial_i\omega^{(1)} \partial^i\omega^{(1)}\, .
\eea
We devote the rest of this section to illustrate how to find such a quantity. 

Let us consider the first two terms in Eq. (\ref{R}) and how they change 
under the time coordinate shift
of Eq. (\ref{timechange}).
 According to our perturbative expansion
\bea
\psi+{\cal H} \frac{\delta \varphi}{\varphi'_0}= {\cal R}^{(1)}+\frac{1}{2} \left[ 
{\mathcal{H}}\frac{\delta^{(2)}
\varphi}{{\varphi'_0}}+ \psi^{(2)} \right]\, ,
\eea
and using Eqs. (\ref{T1}) and (\ref{T2}) this quantity transforms as
\bea \label{T3}
\widetilde{\psi}+{\cal H} \frac{\widetilde{\delta\varphi}}{\varphi'_0}&=&
\psi+{\cal H} \frac{\delta \varphi}{\varphi'_0} 
+\xi^0_{(1)} T-\frac{1}{2} \left( \xi^0_{(1)} \right)^2\left[
 {\cal H}'+2{\cal H}^2
-\frac{{\cal H}}{\varphi'_0}\varphi''_0 \right] 
-\frac{1}{6}\left( 2 \partial^i\omega^{(1)} - \partial^i \xi^{0}_{(1)} 
\right)\partial_i \xi^{0}_{(1)}  \, , 
\eea
where we have set $T={\psi^{(1)}}^{\prime}+2{\cal H} \psi^{(1)}+ {\cal H}{\deu\varphi}^{\prime}/
\varphi'_0$.
Notice that
\be \label{T4}
\widetilde{T}=T+\xi^0_{(1)}\left[ -({\cal H}'+2{\cal H}^2)+\frac{{\cal H}}{\varphi'_0}\varphi''_0 
\right]\, ,
\ee
since the usual first-order transformations  for $\psi^{(1)}$ and 
$\deu \varphi$ are $\widetilde{{\psi}^{(1)}}=\psi^{(1)}
-{\cal H}\, \xi^0_{(1)}$ and $\widetilde{\deu\varphi}=\deu 
\varphi+\varphi'_0\,\xi^0_{(1)}$, respectively. Thus Eq. (\ref{T3}) 
can be written as
\be
\widetilde{\psi}+{\cal H} \frac{\widetilde{\delta \varphi}}{\varphi'_0}=
\psi+{\cal H} \frac{\delta \varphi}{\varphi'_0}+\frac{1}{2} \left( T+\widetilde{T} \right) 
\xi^0_{(1)}
-\frac{1}{6}\left( 2 \partial^i\omega^{(1)} - \partial^i \xi^{0}_{(1)} 
\right)\partial_i \xi^{0}_{(1)} \, .
\ee   
Solving Eq.~(\ref{T4}) for $\xi^0_{(1)}$ and using the 
the first-order transformation 
$\widetilde{\omega^{(1)}}= \omega^{(1)}-\xi^0_{(1)}$ to express 
$\partial_i\xi^0_{(1)}$ and $\partial^i\xi^0_{(1)}$ we finally find
\be
\widetilde{\psi}+{\cal H} \frac{\widetilde{\delta \varphi}}{\varphi'_0}+\frac{1}{2}\, \frac{{\tilde{T}}^{2}}
{{\cal H}'+2{\cal H}^2 -{\cal H}\varphi'_0/\varphi''_0}
-\frac{1}{6} \widetilde{\partial_i\omega^{(1)}}
\widetilde{\partial^i\omega^{(1)}}= 
\psi+{\cal H} \frac{\delta \varphi}{\varphi'_0}+\frac{1}{2}\frac{T^2}{
{\cal H}'+2{\cal H}^2 -{\cal H}\varphi'_0/\varphi''_0}
-\frac{1}{6} \partial_i\omega^{(1)} \partial^i\omega^{(1)}\, ,
\ee       
which shows that ${\cal R}$ -- the combination on the r.h.s. of this equation 
-- 
is indeed gauge-independent.
\\
Notice that, by replacing $\varphi$ with the energy density $\rho$, we can find in an analogous way a 
gauge-invariant expression for the curvature perturbation 
on uniform-density hypersurfaces $\zeta$, which to first order is given by \cite{bst} $- \zeta^{(1)}=
\psi^{(1)}+{\cal H}
\deu \rho/\rho'_0$, where $\rho_0$ is the background energy density. Thus 
at second order 
${\zeta} = {\zeta}^{(1)}+\frac{1}{2} \zeta^{(2)}$ is 
\bea \label{zeta}
- \zeta = - {\zeta^{(1)}}+\frac{1}{2} \left[ {\mathcal{H}}\frac{\delta^{(2)}
\rho} {{\rho'_0}}+ \psi^{(2)} \right] 
+ \frac{1}{2} \frac{ \left( {\psi^{(1)}}^{\prime}+2{\mathcal{H}}\psi^{(1)}
+{\mathcal{H}} {\deu\rho}^{\prime}/ \rho'_0 \right)^2}{{\mathcal{H}}'
+2{\mathcal{H}}^2-{\mathcal{H}}\, \rho''_0/\rho'_0}
-\frac{1}{6} \partial_i\omega^{(1)} \partial^i\omega^{(1)}\, .
\eea  

\section{Einstein Equations}   
\label{EE}
In this section we shall derive the behaviour on large scales
of the comoving curvature perturbation at second order ${\cal R}^{(2)}$
 introduced in Eq.~(\ref{R}). 
Our starting point is to calculate the 
perturbed Einstein equations $\delta^{(2)}G^\mu_{~\nu}=
(\kappa^2/2) \, \delta^{(2)}T^\mu_{~\nu}$ 
in a generalized 
longitudinal gauge. Here $\kappa^2=8\pi \,G_{\rm N}$ and $T^\mu_{~\nu}$ is the
energy-momentum tensor of the inflaton field. From the Einstein equations in this gauge we shall 
pick up an equation for a single unknown function -- the potential $\phi^{(2)}$ --
in a way similar to the procedure used in Ref.~\cite{mfb} to
isolate the equation of motion for $\phi^{(1)}$ in the longitudinal gauge \footnote{We recall that the 
equation of motion for the potential $\phi^{(1)}$ in the longitudinal gauge 
is \cite{mfb} \\
${{\phi}^{(1)}}^{\prime\prime}-\partial_i\partial^i \phi^{(1)}
+ 2\left({\cal H}-
\frac{\varphi_0^{\prime\prime}}{\varphi_0^\prime}\right)
{\phi^{(1)}}^\prime
+2\left({\cal H}^\prime 
-\frac{\varphi_0^{\prime\prime}}{\varphi_0^\prime}{\cal H}\right)
\phi^{(1)}= 0. $\label{footnotesufi1}}.
As in Ref.~\cite{mfb}, but at the second-order level in the perturbations, 
an equation linking $\ded \varphi$ and $\phi^{(2)}$ holds, so that 
it is possible 
to obtain an explicit expression for $\ded \varphi$ and hence for the curvature ${\cal R}^{(2)}$, once the 
equation for $\phi^{(2)}$ has been solved.
Indeed there are many differences with respect to the first-order case, as it will be evident to the reader 
from the details of the calculations which follow. 
The main difficulties arise due to the fact that -- contrary to 
the first-order perturbation theory -- also vector and tensor contributions are present, and to the 
fact that the two scalar potential $\phi^{(2)}$ and $\psi^{(2)}$ differ even in the longitudinal 
gauge for the presence of source terms 
which are quadratic in the first-order perturbations.  

\subsection{Einstein Equations in the generalized longitudinal gauge}
Up to now we have not choosen any particular gauge. 
Hereafter we will work in a 
generalized longitudinal gauge defined as
\bea \label{metric3}
g_{00}&=&-a^2(\tau)(1+2 \phi^{(1)}+\phi^{(2)})\, ,\nonumber\\
g_{0i}&=&0\, ,  \nonumber\\
g_{ij}&=&a^2(\tau)\left[
(1 -2 \psi^{(1)} - \psi^{(2)})\delta_{ij} + \frac{1}{2}\,
\left(\partial_i \chi^{(2)}_j+\partial_j \chi^{(2)}_i+
\chi^{(2)}_{ij}\right)\right]\,.
\eea
One can obtain the Einstein equations in this gauge either by using directly 
the metric tensor in Eq. (\ref{metric3}) or by using the more general metric 
of Eq. (\ref{metric2}), where no gauge choice 
has been specified yet, and reduce the equations
to the longitudinal gauge only at the end. We have performed both 
the computations to have a cross check for the equations obtained. 
In Appendix \ref{A} we give the 
expression for the Einstein tensor in the more general form using the 
metric (\ref{metric2}).\\
We shall now give the Einstein equations in the longitudinal gauge 
at first and second order in the perturbations, respectively
\bea 
\delta^{(1)}G^0_{~0}=\kappa^2 \delta^{(1)}T^0_{~0} & & \textrm{implies}\\
&& \nonumber \\
&6&\,\Ab \phi^{(1)} \,+\,6\,\Aa{\psi^{(1)}}^{\prime}-2\La {\psi^{(1)}} =  
\kappa^2 \left( \phi^{(1)}\,{{\varphi_0}^{\prime}}^2 -
  \deu{{\varphi}^{\prime}}\,{\varphi_0}^{\prime}-\deu {\varphi}\, \frac{\partial V}
{\partial\varphi}\,a^2 \right) \label{001}\\
\delta^{(1)}G^0_{~i}=\kappa^2 \delta^{(1)}T^0_{~i} & & \textrm{implies}\\
&& \nonumber \\
&-&\,2\,\Aa \partial_i \phi^{(1)} \,-\, 2\,\partial_i
 {\psi^{(1)}}^{\prime} = - \kappa^2 \,{\varphi_0}^{\prime} \,\partial^i \deu{{\varphi}} \label{0i1}\\
\delta^{(1)}G^i_{~j}=\kappa^2 \delta^{(1)}T^i_{~j} & & \textrm{implies}\\
&& \nonumber  \\
& &\left(  2\Aa {\phi^{(1)}}^{\prime} \,+\, 4\,\Ac {\phi^{(1)}} \,-\,
2\,\Ab {\phi^{(1)}} \,+\, \La {\phi^{(1)}} \,+\, 4\,\Aa {\psi^{(1)}}^{\prime} \,
+\, 2\,{\psi^{(1)}}^{\prime\prime} 
-  \La {\psi^{(1)}} \right)\delta^{i}_{~j} \\
&-&\, \partial^i\partial_j {\phi^{(1)}} \,+\, \partial^i\partial_j {\psi^{(1)}} = \kappa^2
\left(-\, {\phi^{(1)}}\,{{\varphi_0}^{\prime}}^2 \,+\, \deu{{\varphi}^{\prime}}\,
{\varphi_0}^{\prime} \,-\, \deu{\varphi}\, \frac{\partial V}{\partial {\varphi}}
\,a^2 \right) \delta^i_{~j} \nonumber
\eea
\bea
\label{zero}
\delta^{(2)}G^0_{~0}=\frac{\kappa^2}{2} \delta^{(2)}T^0_{~0} & & \textrm{implies}\\
&& \nonumber \\
&3& \frac{a^{\prime}}{a}\psi^{(2)\prime} \,-
\,\partial_i\,\partial^i\,\psi^{(2)} \,+\, \Ab \phi^{(2)} +\,\Ac\,\phi^{(2)} \,-\,
12 \Ab \left( {\psi^{(1)}} \right)^2\,- \,3\left( {\psi^{(1)}}^{\prime} \right)^2\, \label{34}\\
&-& 8\,\psi^{(1)}\partial_i\partial^i\psi^{(1)}\, -\,3\,\partial_i\psi^{(1)}
\partial^i \psi^{(1)} \, \nonumber\\
&=&\kappa^2 \Big (-\,\frac{1}{2}\, {\ded\varphi}^{\prime}\,\varphi_0^{\prime}  \,-\,
\frac{1}{2}\, {\ded\varphi} \,\frac{\partial V}{\partial \varphi} \,a^2 \,-\,
\frac{1}{2}\,\left( {\deu\varphi}^{\prime} \right)^2 \,-\, \frac{1}{2}
\partial^i {\deu\varphi} \,\partial_i \,{\deu\varphi} \nonumber \\
&-& \frac{1}{2}\,\left( {\deu\varphi} \right)^2 \,\frac{\partial^2
V}{\partial \varphi^2} \,a^2 \,-\, 2\,\left( \psi^{(1)} \right )^2\,{\varphi_0^{\prime}}^2 \,+\,
2\,\psi^{(1)}\,{\deu\varphi}^{\prime}\,\varphi_0^{\prime} \Big)\, ,\nonumber \\
\label{izero}
\delta^{(2)}G^i_{~0}=\frac{\kappa^2}{2} \delta^{(2)}T^i_{~0} & & \textrm{implies}\\
&& \nonumber\\
& &\partial^i\, {\psi^{(2)}}^{\prime}+ 
\,\frac{a^{\prime}}{a}\,\partial^i\,\phi^{(2)} \,+\, \frac{1}{4}
\partial_k\partial^k\left( {\chi^{i(2)}} \right)^{\prime}\,
+\,2\,{\psi^{(1)}}^{\prime}\partial^i \psi^{(1)} \,+\,8\,\psi^{(1)}\partial^i
{\psi^{(1)}}^{\prime} \\
& & =\kappa^2 \Big(\frac{1}{2}\,\,  \varphi_0^{\prime}\,\partial^i \,{\ded\varphi} \,+\,
\partial^i \,{\deu\varphi} \,{{\deu\varphi}}^{\prime} \,+\,
2\,\varphi_0^{\prime}\, \psi^{(1)} \,\partial^i \,{\deu\varphi} \Big)\, , \nonumber \\
\label{ijd}
\delta^{(2)}G^i_{~j}=\frac{\kappa^2}{2} \delta^{(2)}T^i_{~j} & & \textrm{implies}\\
& & \Bigg(\frac{1}{2}\,\LA \phi^{(2)} \,+\, \Aa {\phi^{(2)}}^{\prime}
\,+\, \Ac \phi^{(2)} \,+\, \Ab \phi^{(2)} \,-\, \frac{1}{2}\,\LA \psi^{(2)} \,+\, 2\,\Aa
{\psi^{(2)}}^{\prime} 
+ {\psi^{(2)}}^{\prime\prime} \label{38}\\
&-&8\,\Ac \left( \psi^{(1)} \right)^2 \,+\, 4\,\Ab \left( \psi^{(1)} \right)^2 
\,-\, 8\,\Aa \psi^{(1)}{\psi^{(1)}}^{\prime}
\,-\,3\,\partial_k \psi^{(1)}\,\partial^k \psi^{(1)} \, -\, 4
\,\psi^{(1)}\,\LA \psi^{(1)} \,\nonumber \\
&-&\, \left( {\psi^{(1)}}^{\prime} \right)^2 \Bigg)\,\delta^i_{~j}\,-\frac{1}{2}\,\partial^i\,
\partial_j\,\phi^{(2)} \,+\,
\frac{1}{2}\,\partial^i\partial_j \,\psi^{(2)} \,+\, \frac{1}{2}\,\Aa\left(
\partial_j {\chi^{i(2)}}^{\prime} \,+\, \partial^i
{\chi^{(2)}_j}^{\prime} \,+\, {\chi^{i(2)}_{~j}}^{\prime}\right) \nonumber \\
&+& \frac{1}{4}\,\left( \partial_j {\chi^{i(2)}}^{\prime\prime} \,+\,
\partial^i {\chi^{(2)}_j}^{\prime\prime} \,+\, {\chi^{i(2)}_{~j}}^{\prime\prime}\right)
\,-\, \frac{1}{4}\,\LA \chi^{i(2)}_{~j} \,+\, 2\,\partial^i \psi^{(1)}\,\partial_j
\psi^{(1)} \,+\, 4\,\psi^{(1)}\,\partial^i\partial_j \psi^{(1)}\nonumber \\
&=& \kappa^2 \Big(\frac{1}{2}\, {\ded\varphi}^{\prime}\,\varphi_0^{\prime}
\,-\, \frac{1}{2}\, {\ded\varphi}\,\frac{\partial V }{\partial \varphi}\,a^2 \,+\,
\frac{1}{2}\,\left( {\deu\varphi}^{\prime} \right)^2 \,-\, \frac{1}{2}\,\partial_k
\,{\deu\varphi}\,\partial^k {\deu\varphi} \,+\, 2\,\left( \psi^{(1)} \right)^2\,
{\varphi_0^{\prime}}^2 \nonumber \\
&-& \frac{1}{2}\,\left( \deu\varphi \right)^2 \,\frac{\partial^2 V}{\partial
\varphi^2}\,a^2  \,-\, 2\,\psi^{(1)}\,{{\deu\varphi}}^{\prime}
\,\varphi_0^{\prime}\Big)\,\delta^i_{~j}+\,\,\kappa^2 \left(\partial^i
{\deu\varphi} \,\partial_j {\deu\varphi}\right) \nonumber \,.
\eea
In writing the perturbed Einstein equations at second order we have set 
throughout $\phi^{(1)}=\psi^{(1)}$,
since in the longitudinal gauge at first-order the two scalar 
potentials are equal in the case of a scalar field, and to obtain
 Eqs. (\ref{34}) and (\ref{38}) we have made use of the background relations
$a^{\prime\prime}/a={\cal H}^2+{\cal H}^{\prime}$ and 
$(\kappa^2/2)\,\, {\varphi_0^{\prime}}^2={\cal H}^2-{\cal H}^{\prime}$.\\ 
We shall now describe how to isolate the equation for the potential $\phi^{(2)}$.
We use  
the $(0-0)$-component of Einstein equations, 
the divergence of the $(i-0)$-component and the trace of the 
$(i-j)$-component, 
both performed with the background metric $\delta_{ij}$.  
Notice that the divergence and the trace operations make  the vector and the tensor 
modes  disappear from the equations.  
Thus, we are left with three equations in the three unknown functions, 
$\phi^{(2)}, \psi^{(2)}$, and $\delta^{(2)}\varphi$.\\ 
From the divergence of the $(i-0)$-component 
of Einstein equations it is possible to recover 
an expression for $\delta^{(2)}\varphi$ 
\begin{equation}
\label{defidue} \frac{1}{2}\delta^{(2)}\varphi
\,=\,\frac{\left( \psi^{(2)\prime} \,+\,
{\mathcal{H}}\,\phi^{(2)}  \,+\, \triangle^{-1}\alpha \right)}
{\kappa^2\,\varphi_0^{\prime}}
\,-\,\frac{\triangle^{-1} \beta}{\varphi_0^{\prime}}\, , 
\nonumber 
\end{equation}
where 
\begin{eqnarray}
\alpha&=&2{\psi^{(1)}}^{\prime}\partial_i \,\partial^i \psi^{(1)} \,
+\, 10\, \partial_i {\psi^{(1)}}^{\prime} \partial^i \psi^{(1)}
+\, 8 \psi^{(1)}\, \partial_i \,\partial^i {\psi^{(1)}}^{\prime} 
\label{alpha}\, ,\\ 
\beta&=&\partial_i \,\partial^i 
\delta^{(1)} \varphi \,{\delta^{(1)} \varphi}^{\prime} \,+\,
\partial^i \delta^{(1)} \varphi\,\partial_i {\delta^{(1)}\varphi}^{\prime} 
+\, 2\,\psi^{(1)}\,\partial_i \,\partial^i \delta^{(1)}\varphi
\,\varphi_0^{\prime} \,+\, 2\,\partial_i\psi^{(1)} \,\partial^i \delta^{(1)}
\varphi\,\varphi_0^{\prime} \label{beta}\, ,
\end{eqnarray} 
and $\triangle^{-1}$ is the inverse of the Laplacian operator for the
three spatial-coordinates.
The expression (\ref{defidue}) and its derivative with respect to the
conformal time $\tau$ can be used in the trace of the $(i-j)$ 
equation to obtain
\bea \label{traccia}
\frac{1}{3} \La \phi^{(2)}-\frac{1}{3} \La \psi^{(2)}&=&
8\,\frac{a^{\prime\prime}}{a} \,\, \left( \psi^{(1)} \right)^2  \,-\,
4\, \left( \frac{a^{\prime}}{a} \right)^2\,  \left( \psi^{(1)} \right)^2 \,
+\, 8 \frac{a^{\prime}}{a}\,
\psi^{(1)}{\psi^{(1)}}^{\prime}
+\, \frac{7}{3}\,\partial_i
\psi^{(1)}\,\partial^i \psi^{(1)} \,+\,
\frac{8}{3} \, \psi^{(1)}\,\partial_i \,\partial^i\psi^{(1)}
 \, \nonumber \\
&+&\,  \left( {\psi^{(1)}}^{\prime} \right)^2 
+ \,\triangle^{-1} \alpha^{\prime} \,+\,
2\, \frac{a^{\prime}}{a} \, \triangle^{-1}\alpha \, -\kappa^2 \, \triangle^{-1}\beta^{\prime} 
-\, 2\frac{a^{\prime}}{a}\,\kappa^2\,
\triangle^{-1}\beta 
 + \kappa^2 \bigg[ \frac{1}{2}\,\left( {\delta^{(1)}\varphi}^{\prime} \right)^2 \nonumber \\
&-& \frac{1}{6}\partial_i \delta^{(1)} \varphi\, \partial^i 
\delta^{(1)}\varphi
+\, 2 \left( \psi^{(1)} \right)^2 {\varphi_0^{\prime}}^2 
 -\frac{1}{2}\left( {\delta^{(1)}\varphi} \right)^2 \frac{\partial^2 V}{\partial
\varphi^2}\,a^2 -\, 2 \psi^{(1)} {{\delta^{(1)}\varphi}}^{\prime}
\varphi_0^{\prime} \bigg]\, . 
\eea
Thus a relation between $\psi^{(2)}$ and $\phi^{(2)}$ follows
from Eq.~(\ref{traccia})
\begin{equation} \label{ST}
\psi^{(2)}=\phi^{(2)}-\triangle^{-1} \gamma,
\end{equation}
where $\gamma$ stands for three times the R.H.S  of Eq.~(\ref{traccia}).
Eq. (\ref{ST}) shows that the two scalar potentials 
$\psi^{(2)}$ and $\phi^{(2)}$ differ for quadratic terms 
in the first-order perturbations, as anticipated above. \\
Using Eq. (\ref{ST}) we are now in the position 
to express the other two unknown functions $\psi^{(2)}$ 
and $\ded \varphi$ in terms solely of $\phi^{(2)}$.
From Eq. (\ref{defidue}) we finally obtain

\begin{eqnarray} \label{deltafidue}
\frac{1}{2}\delta^{(2)}\varphi \, = 
\,\frac{\,\left(
\phi^{(2) \prime} \,+\, {\mathcal{H}}\,\phi^{(2)} \,+\, 
\triangle^{-1}\alpha \right)}{\kappa^2\,\varphi_0^{\prime}}
-\frac{\triangle^{-1} \beta}{\varphi_0^{\prime}}
-\frac{\triangle^{-1}\gamma^{\prime}}{\kappa^2\,\varphi_0^{\prime}}\, .
\end{eqnarray}
Plugging Eqs. (\ref{ST}) and (\ref{deltafidue}) into the $(0-0)$ Einstein equation, the  
equation of motion for $\phi^{(2)}$ is derived 
\bea
\label{Long} 
&{{\phi}^{(2)}}&^{\prime\prime}-\partial_i\partial^i \phi^{(2)}
+ 2\left({\cal H}-
\frac{\varphi_0^{\prime\prime}}{\varphi_0^\prime}\right)
{\phi^{(2)}}^\prime
+2\left({\cal H}^\prime 
-\frac{\varphi_0^{\prime\prime}}{\varphi_0^\prime}{\cal H}\right)
\phi^{(2)}=\\
&12&{\cal H}^2\left(\psi^{(1)}\right)^2+3\left({\psi^{(1)}}^\prime
\right)^{2}
+8\psi^{(1)}\partial_i\partial^i\psi^{(1)}+3 \partial_i\psi^{(1)}
\partial^i\psi^{(1)} 
+2\left({\cal H}
+ \frac{\varphi_0^{\prime\prime}}{\varphi_0^\prime}\right)\triangle^{-1}\alpha 
-\triangle^{-1}\alpha^\prime \nonumber \\
&-&2\kappa^2 \left({\cal H}+
\frac{{\varphi_0}^{\prime\prime}}{\varphi_0^\prime} \right)\triangle^{-1}\beta 
+\kappa^2\triangle^{-1}\beta^\prime -\gamma 
+\left( {\cal H} 
-2 \frac{\varphi_0^{\prime\prime}}{\varphi_0^\prime} \right)\triangle^{-1}\gamma^\prime
+\triangle^{-1}\gamma^{\prime\prime}
\nonumber \\
&+&\kappa^2 \left(-\frac{1}{2}\,\left( {\delta^{(1)}\varphi}^{\prime} \right)^2-
\frac{1}{2}\partial_i \delta^{(1)}\varphi\partial^i 
\delta^{(1)}\varphi -\, 2 \left( \psi^{(1)}\right)^2 \varphi_0^{\prime 2} 
-\frac{1}{2}\left( {\delta^{(1)}\varphi} \right)^2 \frac{\partial^2 V}{\partial
\varphi^2}\,a^2 +\, 2 \psi^{(1)} {{\delta^{(1)}\varphi}}^{\prime}
\varphi_0^{\prime} \right)\, .
\nonumber
\eea
Eq. (\ref{Long}) is our master equation. Before solving it, let us stress two important points.  
First, no approximation has been made up to now. In particular notice that this equation  
is exact at any order in the expansion in terms of  the slow-roll parameters.
Secondly, the L.H.S. of Eq. (\ref{Long}) is exactly the same as in 
the equation for 
$\phi^{(1)}$ in the longitudinal gauge at first order (see the footnote \ref{footnotesufi1}). 
However, at second order, the key point is that
Eq. (\ref{Long}) for $\phi^{(2)}$ is not homogeneous, but there is a source made up of terms which 
are quadratic in the first-order perturbations.
Notice that, as in the first-order calculation, it is not necessary to use the perturbed Klein-Gordon 
equation to close the system of the evolution equations for the fluctuactions.
Nevertheless, we also calculated 
the Klein-Gordon equation at second order in the inflaton and metric 
fluctuactions; this is reported in Appendix \ref{B}. 

\subsection{The large-scale curvature perturbation ${\cal R}^{(2)}$ 
in the slow-roll approximation}
We shall now solve Eq.~(\ref{Long}) in order to obtain the expression for the comoving 
curvature perturbation ${\cal R}^{(2)}$ defined in Eq.~(\ref{R}). 
First we rewrite Eq.~(\ref{Long}) in cosmic time $dt=a\, d\tau$, since it is more easy in this way to recognize the 
slow-roll parameters $\epsilon=-\dot{H}/H^2$ and $\eta = 
\epsilon-\left(\ddot\varphi_0/H\dot\varphi_0\right)$ \cite{lr}.   
During a period of inflation $\epsilon$ and $\eta$ must be $\ll 1$, but  
only at a certain point we will perform an expansion to lowest-order in the slow-roll parameters. 
Moreover, where possible, we shall neglect some terms which give a subdominant contribution on large scales. 

Using cosmic time, Eq.~(\ref{Long}) becomes
\bea \label{longt}
&\ddot{\phi}^{(2)}&+H \left( 1- 2 \frac{\ddot{\varphi_0}}{H\dot{
\varphi_0}} \right) \dot{\phi}^{(2)}+2 H^2 \left( \frac{\dot{H}}{H^2} -\frac{\ddot{\varphi_0}}{H\dot{
\varphi_0}} \right) \phi^{(2)}-\frac{1}{a^2}\La \phi^{(2)}= \\
&-&24 H^2 \left( 1+\frac{\dot{H}}{H^2} \right) \left( \psi^{(1)} \right)^2-24 H \psi^{(1)} 
\dot{\psi}^{(1)}-\frac{4}{a^2}\partial_i \psi^{(1)} 
\partial^i \psi^{(1)}- \frac{2}{a} H \left( 1-\frac{\ddot{\varphi_0}}{H\dot{
\varphi_0}} \right)\triangle^{-1}\alpha \nonumber \\
&-&\frac{4}{a} \triangle^{-1}\dot{\alpha}
+2 \frac{\kappa^2}{a} H \left( 1-\frac{\ddot{\varphi_0}}{H\dot{
\varphi_0}} \right)\triangle^{-1}\beta +4\frac{\kappa^2}{a} \triangle^{-1}\dot{\beta}-2  
\frac{\ddot{\varphi_0}}{\dot{\varphi_0}}\triangle^{-1}\dot{\gamma}+\triangle^{-1}\ddot{\gamma}\nonumber \\
&-& \kappa^2 \left( 2 \left( {\deu \varphi}^{\displaystyle{\cdot}} \right)^2 +8 {\dot{\varphi_0}}^2 
\left( \psi^{(1)} \right)^2-
\frac{\partial^2 V}{\partial\varphi^2}\left( \deu \varphi \right)^2-8\, \dot{\varphi_0} \psi^{(1)}
 \, \left( \delta^{(1)} \varphi \right)^{\displaystyle{\cdot}} \right)\, ,   \nonumber
\eea
where we have replaced 
the term $-\gamma$ appearing in Eq.~(\ref{Long}) with its explicit definition given by Eqs.~(\ref{traccia}) 
and (\ref{ST}). 
Notice that in the source on the R.H.S. of Eq.~(\ref{longt}) there appears always the combination 
$\Delta^{-1}\left(\kappa^2 \beta/a-\alpha/a\right)$ 
and its derivative with respect to cosmic time. Let us now calculate 
such a combination. The quantity $\alpha$ defined in Eq.~(\ref{alpha}) can be rewritten as
\be
\frac{\alpha}{a}=2 \La \left( \psi^{(1)} \dot{\psi}^{(1)} \right)+6 \left( \partial_i \psi^{(1)} 
\partial^i \dot{\psi}^{(1)}+\psi^{(1)} \La \dot{\psi}^{(1)} \right) \, .
\ee
Using the equation of motion~(\ref{001}) to express 
$\left( \deu \varphi \right)^{\displaystyle{\cdot}}$, the quantity $\beta$, 
Eq.~(\ref{beta}), turns out to be
\bea
\frac{\beta}{a}&=&\frac{1}{2} \frac{\ddot{\varphi_0}}{\dot{\varphi_0}} \La \left( \deu \varphi \right)^2
+3 \dot{\varphi_0}\, 
\partial_i \psi^{(1)} \partial^i \deu \varphi+3 \dot{\varphi_0}\,  \psi^{(1)} \La \deu \varphi \nonumber \\ 
&+& \frac{2}{\kappa ^2
\dot{\varphi_0}} \partial_k \deu \varphi \, \partial^k \left( \La \psi^{(1)} \right) 
+ \frac{2}{\kappa ^2\dot{\varphi_0}} \La \deu \varphi\,  \LA \deu \psi^{(1)} \, .
\eea 
Using the equation of motion for $\psi^{(1)}$ in the longitudinal gauge
\be \label{psieq}
\dot{\psi}^{(1)}+H \psi^{(1)}=\frac{\kappa^2}{2} \dot{\varphi_0} \deu \varphi\, ,
\ee
which can be derived from Eq.~(\ref{0i1}) with $\phi^{(1)}=\psi^{(1)}$, we get
\bea
\kappa^2 \frac{\beta}{a}-\frac{\alpha}{a}&=&\frac{\kappa^2}{2} 
\frac{\ddot{\varphi_0}}{\dot{\varphi_0}} \La \left( \deu \varphi \right)^2+3 H \La \left( \psi^{(1)} \right)^2
-2 \La \left( \psi^{(1)} \dot{\psi}^{(1)} \right) \nonumber \\ 
&+&\frac{2}{a^2 \dot{\varphi_0}} 
\partial_k \deu \varphi \, \partial^k \left( \La \psi^{(1)} \right) 
+ \frac{2}{a^2 \dot{\varphi_0}} \La \deu \varphi\,  \LA \deu \psi^{(1)}\, ,  
\eea
and therefore 
\bea \label{combination}
\Delta^{-1} \left( \kappa^2 \frac{\beta}{a}-\frac{\alpha}{a} \right) &=&
 \frac{\kappa^2}{2} 
\frac{\ddot{\varphi_0}}{\dot{\varphi_0}} \left( \deu \varphi \right)^2+3 H \left( \psi^{(1)} \right)^2
-2 \left( \psi^{(1)} \dot{\psi}^{(1)} \right) \nonumber \\ 
&+&\frac{2}{\dot{\varphi_0}} \Delta^{-1} \left[  
\partial_k \deu \varphi \, \partial^k \left( \La \psi^{(1)} \right) 
+ \La \deu \varphi\,  \LA  \psi^{(1)} \right]\, .
\eea
The master equation~(\ref{longt}) can be simplified if we drop those terms which are next to leading-order 
in the slow-roll parameters
\bea 
&\ddot{\phi}^{(2)}&+H \dot{\phi}^{(2)}-\frac{1}{a^2}\La \phi^{(2)}= \\
&-&  12 \kappa^2 H \deu \varphi\,  \psi^{(1)}\dot{\varphi_0} + 
3 \kappa^2 H \frac{\ddot{\varphi_0}}{\dot{\varphi_0}} \left( \deu \varphi \right)^2 
-12 H \psi^{(1)} \dot{\psi}^{(1)} +18 H^2 \left( \psi^{(1)} \right)^2\nonumber \\
&+&4  \left[ \frac{\kappa^2}{2} 
\frac{\ddot{\varphi_0}}{\dot{\varphi_0}} \left( \deu \varphi \right)^2+3 H \left( \psi^{(1)} \right)^2
-2 \left( \psi^{(1)} \dot{\psi}^{(1)} \right) \right]^{\displaystyle{\cdot}}
-2 \frac{\ddot{\varphi_0}}{\dot{\varphi_0}}\triangle^{-1}\dot{\gamma}+\triangle^{-1}\ddot{\gamma}\nonumber \\
&+& \frac{12 H}{a^2 \dot{\varphi_0}} \Delta^{-1} 
\left[ \partial_k \deu \varphi \, \partial^k \left( \La \psi^{(1)} \right) 
+ \La \deu \varphi\,  \LA  \psi^{(1)} \right] \nonumber \\
&+& \frac{8}{a^2 \dot{\varphi_0}} \Delta^{-1} \left[  
\partial_k \deu \varphi \, \partial^k \left( \La \psi^{(1)} \right) 
+ \La \deu \varphi\,  \LA  \psi^{(1)} \right]^{\displaystyle{\cdot}}
- \frac{4}{a^2} \partial_i \psi^{(1)} \partial^i \psi^{(1)}
 +\kappa^2 \frac{\partial^2 V}{\partial\varphi^2} \left( \deu \varphi \right)^2 
\nonumber \\ 
&-&\frac{8}{\kappa^2 \dot{\varphi_0}^2} \left( \frac{\La \psi^{(1)}}{a^2} \right)^2
+\frac{8}{a^2} \psi^{(1)} \La \psi^{(1)} -\frac{8}{a^2} 
\frac{\ddot{\varphi_0}}{\dot{\varphi_0}} \frac{\deu \varphi}{\dot{\varphi_0}} \La \psi^{(1)}\, ,  
\eea
where we have used Eq.~(\ref{psieq}) so that $-24 H^2 \left( 1+\dot{H}/H^2 \right) 
\left( \psi^{(1)} \right)^2-24 H \psi^{(1)} \dot{\psi}^{(1)}=-24 \dot{H} 
\left( \psi^{(1)} \right)^2 - 12 \kappa^2 H \deu \varphi\,  
\psi^{(1)}\dot{\varphi_0}$, we have used the expression 
for $\left( \deu \varphi \right)^{\displaystyle{\cdot}}$ 
from Eq.~(\ref{001}), and we also explicitly written the 
combination~(\ref{combination}).

Let us notice that through Eq.~(\ref{psieq}) we find   
\bea
&-& 12 \kappa^2 H \deu \varphi\,  \psi^{(1)}\dot{\varphi_0} + 
3 \kappa^2 H \frac{\ddot{\varphi_0}}{\dot{\varphi_0}} \left( \deu \varphi \right)^2 
-12 H \psi^{(1)} \dot{\psi}^{(1)} +18 H^2 \left( \psi^{(1)} \right)^2= \nonumber \\
&-&18 \kappa^2 H \deu \varphi\,  \psi^{(1)}\dot{\varphi_0}+ 
3 \kappa^2 H \frac{\ddot{\varphi_0}}{\dot{\varphi_0}} \left( \deu \varphi \right)^2 +
30 H^2 \left( \psi^{(1)} \right)^2 = \nonumber \\
&-&12 H \left( \psi^{(1)\, 2}  \right)^{\displaystyle{\cdot}} + 6 \left( \dot{\psi}^{(1)} \right)^2
 - \kappa^2 \frac{\partial^2 V}{\partial\varphi^2} \left( \deu \varphi \right)^2\, ,
\eea
where the last passage is valid to lowest-order in the slow-roll parameters.
Thus we finally obtain
\bea \label{eqf}
&\ddot{\phi}^{(2)}&+H \dot{\phi}^{(2)}+2 H^2 \left( \frac{\dot{H}}{H^2} -\frac{\ddot{\varphi_0}}{H\dot{
\varphi_0}} \right) \phi^{(2)}-\frac{1}{a^2}\La \phi^{(2)}= \nonumber \\
&-&12 H \left( \psi^{(1)\, 2}  \right)^{\displaystyle{\cdot}} + 6 \left( \dot{\psi}^{(1)} \right)^2
+4  \left[ \frac{\kappa^2}{2} 
\frac{\ddot{\varphi_0}}{\dot{\varphi_0}} \left( \deu \varphi \right)^2+3 H \left( \psi^{(1)} \right)^2
-2 \left( \psi^{(1)} \dot{\psi}^{(1)} \right) \right]^{\displaystyle{\cdot}}\nonumber \\
&-&2 \frac{\ddot{\varphi_0}}{\dot{\varphi_0}}\triangle^{-1}\dot{\gamma}+\triangle^{-1}\ddot{\gamma}
+ \frac{12 H}{a^2 \dot{\varphi_0}} \Delta^{-1} 
\left[ \partial_k \deu \varphi \, \partial^k \left( \La \psi^{(1)} \right) 
+ \La \deu \varphi\,  \LA  \psi^{(1)} \right] \nonumber \\
&+& \frac{8}{a^2 \dot{\varphi_0}} \Delta^{-1} \left[  
\partial_k \deu \varphi \, \partial^k \left( \La \psi^{(1)} \right) 
+ \La \deu \varphi\,  \LA  \psi^{(1)} \right]^{\displaystyle{\cdot}}
- \frac{4}{a^2} \partial_i \psi^{(1)} \partial^i \psi^{(1)}\nonumber \\
&-&\frac{8}{\kappa^2 \dot{\varphi_0}^2} \left( \frac{\La \psi^{(1)}}{a^2} \right)^2
+\frac{8}{a^2} \psi^{(1)} \La \psi^{(1)} -\frac{8}{a^2} 
\frac{\ddot{\varphi_0}}{\dot{\varphi_0}} \frac{\deu \varphi}{\dot{\varphi_0}} \La \psi^{(1)} \, .
\eea
 Integrating Eq.~(\ref{eqf}) we find
\bea \label{integration}
\dot{\phi}^{(2)}+H \phi^{(2)}&=&-12 H  \left( \psi^{(1)}  \right)^2+
4  \left[ \frac{\kappa^2}{2} 
\frac{\ddot{\varphi_0}}{\dot{\varphi_0}} \left( \deu \varphi \right)^2+3 H \left( \psi^{(1)} \right)^2
-2 \left( \psi^{(1)} \dot{\psi}^{(1)} \right) \right] 
-  \frac{\ddot{\varphi_0}}{\dot{\varphi_0}}\triangle^{-1} \gamma+\triangle^{-1}\dot{\gamma}\nonumber \\
&+&6 \int \left( \dot{\psi}^{(1)} \right)^2 dt
+ \int  \frac{12 H}{a^2 \dot{\varphi_0}} \Delta^{-1} 
\left[ \partial_k \deu \varphi \, \partial^k \left( \La \psi^{(1)} \right) 
+ \La \deu \varphi\,  \LA  \psi^{(1)} \right] dt \nonumber \\   
&-& 4 \int  \frac{1}{a^2} \left( \partial_i \psi^{(1)} \partial^i \psi^{(1)}  
\right) dt 
- \int \frac{8}{\kappa^2 \dot{\varphi_0}^2} \left( \frac{\La \psi^{(1)}}{a^2} \right)^2 dt
+ \int \frac{8}{a^2} \psi^{(1)} \La \psi^{(1)} dt 
\nonumber \\
&-& \int \frac{8}{a^2} \frac{\ddot{\varphi_0}}{\dot{\varphi_0}} 
\frac{\deu \varphi}{\dot{\varphi_0}} \La \psi^{(1)} dt \, .
\eea

We are now in the position to calculate the second-order comoving curvature 
perturbation ${\cal R}^{(2)}$.
From Eqs.~(\ref{R}),  (\ref{ST}) and (\ref{deltafidue}) we get
\bea \label{formula:R2}
{\cal R}^{(2)} = &2& \frac{H}{\kappa^2 \dot{\varphi_0}^2} \left[ \dot{\phi}^{(2)}+H {\phi}^{(2)}
+\triangle^{-1} \left( \frac{\alpha}{a} \right) \right]
 -2  \frac{H}{\dot{\varphi_0}^2}
 \triangle^{-1} \left( \frac{\beta}{a} \right) -2 
\frac{H}{\kappa^2 \dot{\varphi_0}^2} 
\triangle^{-1}\dot{\gamma}+  {\phi}^{(2)}-\triangle^{-1} \gamma \nonumber \\
&+& \frac{1}{2} \frac{ \left( \dot{\psi^{(1)}}+2 H \psi^{(1)}
+ H \delta^{(1)}\dot{\varphi}/ \dot{\varphi_0} \right)^2}{H^2 \left( 
2+\dot{H}/{H^2}- \ddot{\varphi_0}/H\dot{\varphi_0} \right)} \, ,
\eea
where the last term is the part of ${\cal R}^{(2)}$ in Eq.~(\ref{R}) 
which is quadratic in the first-order 
perturbations. Note that it is of the order of ${\cal O}(\epsilon^2, \eta^2)$
and thus we 
shall neglect it in the following. 
In Eq.~(\ref{formula:R2}) there appears once more the combination 
$\Delta^{-1}\left(\kappa^2 \beta/a-\alpha/a\right)$ . 
Thus using Eq.~(\ref{combination}) and Eq.~(\ref{integration}) we obtain 
\bea \label{R2}
{\cal R}^{(2)}&=&3 H  \frac{\ddot{\varphi_0}}{\dot{\varphi_0}} \left( \frac{\deu \varphi}{\dot{\varphi_0}} \right)^2 
-12 \frac{H}{\kappa^2 \dot{\varphi_0}^2} \psi^{(1)} \dot{\psi}^{(1)} -6 \frac{H^2}{\kappa^2 \dot{\varphi_0}^2} 
\left( \psi^{(1)} \right)^2 
+\frac{12 H}{\kappa^2 \dot{\varphi_0}^2} \int \left( \dot{\psi}^{(1)} \right)^2 dt \nonumber \\
&+&\frac{24 H^2}{\kappa^2 \dot{\varphi_0}^3} \int  \frac{1}{a^2} \Delta^{-1} 
\left[ \partial_k \deu \varphi \, \partial^k \left( \La \psi^{(1)} \right) 
+ \La \deu \varphi\,  \LA  \psi^{(1)} \right] dt 
-\frac{8 H} {\kappa^2 \dot{\varphi_0}^2}\int  \frac{1}{a^2} \left( \partial_i \psi^{(1)} \partial^i \psi^{(1)}  
\right) dt 
\nonumber \\
&-&  \frac{16 H}{\kappa^4 \dot{\varphi_0}^4} \int  \left( \frac{\La \psi^{(1)}}{a^2} \right)^2 dt
+  \frac{16 H}{\kappa^2 \dot{\varphi_0}^2} \int \frac{1}{a^2} \psi^{(1)} \La \psi^{(1)} dt 
-  \frac{16 H}{\kappa^2 \dot{\varphi_0}^3} \frac{\ddot{\varphi_0}}{\dot{\varphi_0}} 
\int \frac{1}{a^2} \deu \varphi \, \La \psi^{(1)} dt
\nonumber \\
&=& -3 \eta H^2 \left( \frac{ \deu \varphi }{\dot{\varphi_0}} \right)^2
       +6 \frac{H^2}{\kappa^2 \dot{\varphi_0}^2} \left( \psi^{(1)} \right)^2
       -6 H \frac{\delta\varphi}{\dot{\varphi_0}} \psi^{(1)}
       +3 \epsilon H^2 \frac{(\delta\varphi)^2}{\dot{\varphi_0}^2}
       +\frac{12H}{\kappa^2 \dot{\varphi_0}^2} \int \left( \dot{\psi} \right)^2 dt\nonumber \\
&+&\frac{24 H^2}{\dot{\varphi_0}^3} \int  \frac{1}{a^2} \Delta^{-1} 
\left[ \partial_k \deu \varphi \, \partial^k \left( \La \psi^{(1)} \right) 
+ \La \deu \varphi\,  \La  \psi^{(1)} \right] dt 
-\frac{8 H} {\kappa^2 \dot{\varphi_0}^2}\int  \frac{1}{a^2} \left( \partial_i \psi^{(1)} \partial^i \psi^{(1)}  
\right) dt \nonumber \\
&-&  \frac{16 H}{\kappa^4 \dot{\varphi_0}^4} \int  \left( \frac{\La \psi^{(1)}}{a^2} \right)^2 dt
+  \frac{16 H}{\kappa^2 \dot{\varphi_0}^2} \int \frac{1}{a^2} \psi^{(1)} \La \psi^{(1)} dt 
-  \frac{16 H}{\kappa^2 \dot{\varphi_0}^3} \frac{\ddot{\varphi_0}}{\dot{\varphi_0}} 
\int \frac{1}{a^2} \deu \varphi \, \La \psi^{(1)} dt \nonumber\\
&-&  2 
\frac{H}{\kappa^2 \dot{\varphi_0}^2}\frac{\ddot{\varphi_0}}{\dot{\varphi_{0}}} 
\triangle^{-1}\gamma-\triangle^{-1} \gamma\, ,
\eea
where the last passage is valid to 
lowest order in the slow-roll parameters and 
we have used the equation of motion~(\ref{psieq}).
In Eq.~(\ref{R2}) the terms which are not integrated in time can be safely 
treated in the long-wavelength limit. 
To lowest order in the slow-roll parameters and on large scales, 
$k \ll aH$, $\psi^{(1)}$ can be considered as constant and 
\bea 
\psi^{(1)}&=&\frac{\kappa^2}{2}\frac{\dot{\varphi_0}}{H} \deu \varphi=\epsilon \, H\frac{\deu \varphi}{\dot{
\varphi_0}}\label{psi1}
\eea 
which can be derived 
from the equations of motion~(\ref{psieq}) and (\ref{KG1L}) 
in the longitudinal gauge 
for $\psi^{(1)}$ and $\deu \varphi$, respectively. 
On the other hand, from the definition of the comoving 
curvature perturbation at first order, Eq. 
(\ref{R1}), and using Eq.~(\ref{psi1}) 
we can write ${\cal{R}}^{(1)}= H \deu \varphi/\dot{
\varphi_0}$ to lowest order in the slow-roll 
parameters and on large scales. Thus in these approximations Eq.~(\ref{psi1})  
can be rewritten as
\bea
\psi^{(1)}&=&\epsilon\, {\cal{R}}^{(1)}\, .
\label{psi1bis} 
\eea 
Performing various integrations by parts in expression (\ref{R2}), 
using the perturbation equations at first-order and properly
substracting the contributions in the far ultraviolet, we arrive at the final
expression for the comoving curvature perturbation
\be
{\cal R}^{(2)}= 
\left(\eta-3\epsilon\right) \left( {\cal R}^{(1)} \right)^2
+{\cal I}\, ,
\ee
where

\bea
{\cal I}&=&-\frac{2}{\epsilon}\,\int \frac{1}{a^2} \psi^{(1)} 
\La \psi^{(1)} dt -\frac{4}{\epsilon}\,
\int  \frac{1}{a^2} 
\left( \partial_i \psi^{(1)} \partial^i \psi^{(1)}  
\right) dt\nonumber\\
&-&\frac{4}{\epsilon}\,\int\,\left(\ddot{\psi}^{(1)}\right)^2 dt
+\left(\epsilon-\eta\right)\triangle^{-1}\partial_i R^{(1)}
\partial^i R^{(1)}\, .
\eea


\section{Discussion and Conclusions}

In this paper we have provided a complete analysis  of the second-order scalar
perturbations during inflation
leading to the derivation of the 
gauge-invariant  comoving
curvature perturbation ${\cal R}$. 

The comoving
curvature perturbation receives a
contribution which is quadratic in ${\cal R}^{(1)}$. The 
total curvature perturbation will then have a
non-Gaussian
$(\chi^2)$ component.  Reminding that the gauge-invariant potential
$\Phi$ (which is related to Bardeen's variable \cite{bardeen} by
$\Phi=-\Phi_H$) and the curvature perturbation $\cal R$ are related by
$\Phi=\frac{3}{5} {\cal R}$, the following simple relation in
configuration space holds

\begin{equation}
\Phi({\bf x}) = \Phi_{\rm Gauss}({\bf x}) + 
\int\, d^3y\,d^3 z\,{\cal K}\left({\bf y},{\bf z}\right)
\Phi_{\rm Gauss}({\bf x}-{\bf y})\Phi_{\rm Gauss}({\bf x}-{\bf z})+
{\rm constant}
\label{gauss}
\end{equation}
which is valid on superhorizon scales and where
the constant is such that $\langle \Phi({\bf x})\rangle=0$.
Here $\Phi_{\rm Gauss}=\frac{3}{5}
{\cal R}^{(1)}$ is a Gaussian random field. The 
non-Gaussianity kernel in momentum space is given by

\be
{\cal K}\left({\bf k}_1,{\bf k}_2\right)=\frac{5}{6}\left(\eta-3\epsilon\right)
+ f_{{\cal K}}\left({\bf k}_1,{\bf k}_2\right)\, ,
\ee
where $f_{{\cal K}}\left({\bf k}_1,{\bf k}_2\right)$ is directly
related to the function ${\cal I}$ and is first-order in the
slow-roll parameters. The gravitational potential bispectrum
then reads

\be
\langle \Phi({\bf k}_1) \Phi({\bf k}_2) \Phi({\bf k}_3)
\rangle=(2\pi)^3\,\delta^{(3)}\left({\bf k}_1+{\bf k}_2+{\bf k}_3\right)
\,\left[2\,{\cal K}\left({\bf k}_1,{\bf k}_2\right)\,
{\cal P}_\Phi({\bf k}_1){\cal P}_\Phi({\bf k}_2)+{\rm cyclic}\right]\, ,
\ee
where ${\cal P}_\Phi({\bf k})$ is the power spectrum of the gravitational
potential.
We can also  define an effective ``momentum-dependent''
parameter $f_{\rm NL}$ given by\footnote{Notice that according to 
our definition
of $\Phi$, our $f_{\rm NL}$ has the opposite sign of that in Ref. \cite{ks}.}

\begin{equation}
\label{pp}
f_{\rm NL} \sim \frac{5}{12} \left(n_S-1 \right)+
f_{{\cal K}}\left({\bf k}_1,{\bf k}_2\right)  \, ,
\end{equation}
where we have made use of the relation between the 
spectral index $n_S=1 - 6 \epsilon + 2 \eta$ for the  
scalar perturbations and the slow-roll parameters.
In the final expression (\ref{pp}) 
we have not written possible terms of the order of 
${\cal O}(\epsilon^2,\eta^2)\Delta N$ which might be sizeable for 
certain classes of inflationary models ($\Delta N$ is the number of 
e-foldings from the time at which a given scale crosses the horizon and the 
end of inflation; for large-scale CMB anisotropies $\Delta N \approx 60$).
Whether these terms cancel out, or equivalently whether the curvature 
perturbation remains constant on super-horizon scales also at next-to-leading 
order in the slow-roll parameters remains an open issue. 
However, if present, terms of order of ${\cal O}(\epsilon^2,\eta^2)\Delta N$ 
resemble those found in Refs.~\cite{FRS,gup},  
coming from the self-interactions of the inflaton field.
The primordial gauge-invariant potential bispectrum leads to a nonzero
CMB bispectrum via the Sachs-Wolfe effect $\left(\Delta T/T\right)_{SW}=
(1/3)\Phi$.

Deviations from a scale-invariant spectrum can make the primordial 
non-Gaussianity non-negligible.
The possible 
presence of non-Gaussianity in primordial cosmological perturbations is only 
mildly constrained by existing observations \cite{Vetal,v2}.
Recent analyses of the angular bispectrum from 4-year COBE
data \cite{ketal} yield a weak upper limit, $\vert f_{\rm NL}\vert < 1.5
\times 10^3$. The analysis of the diagonal angular
bispectrum of the Maxima dataset \cite{setal} also provides a very
weak constraint: $\vert f_{\rm NL} \vert < 2330$.
According to Ref. \cite{ks}, however, the minimum value of
$\vert f_{\rm NL}\vert$ that will become detectable from the analysis of
MAP and {\it Planck} data, after properly subtracting detector noise and
foreground contamination, is about  $20$ and $5$, respectively.
These results imply that detecting non-Gaussianity  at
the level emerging from
our second-order calculation will represent a challenge  for the
forthcoming satellite experiments.\\

\noindent {\bf Note added}: While revising this paper, a related 
work by Maldacena, astro-ph/0210603, appeared, where the
three-point correlation function for the curvature perturbation
is obtained by computing the cubic term contributions to the 
Lagrangian. Our results, where comparison is possible, agree with
those in astro-ph/0210603.

\section*{Acknowledgments}
We wish to thank M.~Bruni for useful discussions on gauge transformations.

\vskip 1cm
\appendix
\setcounter{equation}{0}
\def\theequation{A.\arabic{equation}}
\vskip 0.2cm
\section{Perturbing gravity at second order}
\label{A}
\subsection{Basic notation}
The number of spatial dimensions is $n=3$.
Greek indices ($\alpha, \beta, ..., \mu, \nu, ....$)
 run from 0 to 3, while latin indices ($a,b,...,i,j,k,....
m,n...$) run from 1 to 3. 
The total spacetime metric $g_{\mu \nu}$ has signature ($-,+,+,+$). 
The connection coefficients are defined as
\be
\label{conness} \Gamma^\alpha_{\beta\gamma}\,=\,
\frac{1}{2}\,g^{\alpha\rho}\left( \frac{\partial
g_{\rho\gamma}}{\partial x^{\beta}} \,+\, \frac{\partial
g_{\beta\rho}}{\partial x^{\gamma}} \,-\, \frac{\partial
g_{\beta\gamma}}{\partial x^{\rho}}\right)\, .
\ee
The Riemann tensor is defined as
\be
R^{\alpha}_{~\beta\mu\nu}=
\Gamma^{\alpha}_{\beta\nu,\mu}-\Gamma^{\alpha}_{\beta\mu,\nu}+
\Gamma^{\alpha}_{\lambda\mu}\Gamma^{\lambda}_{\beta\nu}-
\Gamma^{\alpha}_{\lambda\nu}\Gamma^{\lambda}_{\beta\mu} \,.
\ee

The Ricci tensor is a contraction of the Riemann tensor
\be
R_{\mu\nu}=R^{\alpha}_{~\mu\alpha\nu} \,,
\ee
and in terms of the connection coefficient it is given by
\begin{equation}
R_{\mu\nu}\,=\, \partial_\alpha\,\Gamma^\alpha_{\mu\nu} \,-\,
\partial_{\mu}\,\Gamma^\alpha_{\nu\alpha} \,+\,
\Gamma^\alpha_{\sigma\alpha}\,\Gamma^\sigma_{\mu\nu} \,-\,
\Gamma^\alpha_{\sigma\nu} \,\Gamma^\sigma_{\mu\alpha}\,.
\end{equation}
The Ricci scalar is given by contracting the Ricci tensor
\be
R=R^{\mu}_{~\mu} \,.
\ee
The Einstein tensor is defined as
\be
G_{\mu\nu}=R_{\mu\nu}-\frac{1}{2}g_{\mu\nu}R \,.
\ee
The Einstein equations are written as $G_{\mu\nu}=\kappa^2 T_{\mu\nu}$, so that $\kappa^2=8\pi G_{\rm N}$, 
where $G_{\rm N}$ is the usual Newtonian gravitational constant.\\ 
In the following expressions we have chosen a specific ordering of the terms. In the expressions in which
two spatial indices appear, such as Eq.~(\ref{gammaij}), we have assembled together the terms proportional to
 $\delta_{ij}$. The intrinsically second-order terms precede the source terms which are quadratic in 
the first-order perturbations. The second-order fluctuations have been listed in the following order as 
$\phi^{(2)}$, $\psi^{(2)}$, $\omega^{(2)}$, $\omega^{(2)}_i$, $\chi^{(2)}$, $\chi^{(2)}_i$ 
and $\chi^{(2)}_{ij}$, respectively. This ordering simplifies the analogy between the first-order 
and the second-order equations and allows to obtain immediately the expressions in a given gauge.
\subsection{The connection coefficients}
In a spatially flat Robertson-Walker background the connection coefficients are
\begin{equation}
\Gamma^0_{00}\,=\, \Aa \,; \qquad
\Gamma^i_{0j}\,=\,\Aa\,\delta^i_{~j}\,; \qquad
\Gamma^0_{ij}\,=\,\Aa\,\delta_{ij}\,;
\end{equation}
\begin{equation}
\Gamma^i_{00}\quad=\quad\Gamma^0_{0i}\quad=\quad\Gamma^i_{jk}\quad=\quad
0 \,.
\end{equation}
The first-order perturbed connection coefficients corresponding to 
first-order metric perturbations 
in Eq.~(\ref{metric2}) are 
\bea
\deu{\Gamma^0_{00}}&=& {\phi^{(1)}}^{\prime} \, , \\
\deu{\Gamma^0_{0i}}&=& \partial_i\, \phi^{(1)} \,+\,
\frac{a^{\prime}}{a}\partial_i\,\omega^{(1)} \, , \\
\deu{\Gamma^i_{00}} &=& \frac{a^{\prime}}{a}\,\partial^i \omega^{(1)} \,+\,
\partial^i {\omega^{(1)}}^{\prime} \,+\, \partial^i \phi^{(1)} \, , \\
\deu{\Gamma^0_{ij}}&=&
-\,2\,\frac{a^{\prime}}{a}\,\phi^{(1)}\,\delta_{ij} \,-\,
\partial_i \partial_j \omega^{(1)} \,-\,
2\,\frac{a^{\prime}}{a}\,\psi^{(1)}\,\delta_{ij} 
- {\psi^{(1)}}^{\prime}\,\delta_{ij} \,-\,
\frac{a^{\prime}}{a}\,D_{ij} \chi^{(1)} \,+\, \frac{1}{2}\,D_{ij}{\chi^{(1)}}^{\prime}
\, , \\
\deu{\Gamma^i_{0j}}&=& - \,{\psi^{(1)}}^{\prime}\delta_{ij} \,+\,
\frac{1}{2}\,D_{ij}{\chi^{(1)}}^{\prime} \, , \\
\deu{\Gamma^i_{jk}} &=& \partial_j \psi^{(1)} \,\gamma_k^i \,-\,
\partial_k \psi^{(1)}\, \gamma_j^i \,+\, \partial^i \psi^{(1)} \,\gamma_{jk}
\,-\, \frac{a^{\prime}}{a}\,\partial^i \omega^{(1)}
\,\gamma_{jk} 
+ \frac{1}{2}\,\partial_j D^i_k \chi^{(1)} \,+\, \frac{1}{2}\,\partial_k D^i_j \chi^{(1)} \,-\,
\frac{1}{2}\,\partial^i D_{jk} \chi^{(1)} \, .
\eea
At second order we get:
\bea 
\ded{\Gamma^0_{00}}\,&=&\frac{1}{2}\,{\phi^{(2)}}^{\prime}\,-\,
2\,\phi^{(1)}\, {\phi^{(1)}}^{\prime} \,+\, \partial^k \phi^{(1)}\,\partial_k \omega^{(1)} \,
+\,\Aa
\partial^k \omega^{(1)}\,\partial_k \omega^{(1)} \,+\,\partial^k \omega^{(1)}\,\partial_k 
{\omega^{(1)}}^{\prime}\,, \\
\ded{\Gamma^0_{0i}}&=& \frac{1}{2} \partial_i\,\phi^{(2)} \,+\,
\frac{1}{2}\,\Aa (\partial_i \omega^{(2)} + \omega_i^{(2)} ) \,-\, 2\,\phi^{(1)}\,\partial_i\,\phi^{(1)}
\,-\, 2\,\Aa \phi^{(1)}\,\partial_i \omega^{(1)}-\, {\psi^{(1)}}^{\prime}\partial_i \omega^{(1)} \\ 
&+&\frac{1}{2}\,\partial^k \omega^{(1)}\,D_{ik}{\chi^{(1)}}^{\prime} \nonumber  \,, \\
\ded{\Gamma^i_{00}} \,&=& \,
 \frac{1}{2}\,\partial^i\,\phi^{(2)} \,+\, \frac{1}{2}\,\Aa (\partial^i \omega^{(2)} +
 \omega^{i(2)}) \,+\, \frac{1}{2}\,\left( \partial^i {\omega^{(2)}}^{\prime} + 
\left( \omega^{i{(2)}} \right)^{\prime} \right)
\,+\,2\,\psi^{(1)}\,\partial^i\,\phi^{(1)} \,-\, {\phi^{(1)}}^{\prime}\partial^i 
\omega^{(1)} \\
&+&\, 2\,\Aa \psi^{(1)} \,\partial^i \omega^{(1)} \,+\, 2\,\psi^{(1)}\,\partial^i
{\omega^{(1)}}^{\prime} \,-\, \partial_k \phi^{(1)} \,D^{ik}\chi^{(1)} \,-\, 
\Aa \partial_k \omega^{(1)}
\,D^{ik}\chi^{(1)} \,-\, \partial_k {\omega^{(1)}}^{\prime}D^{ik}\chi^{(1)} \nonumber \,, \\
\label{gammaij} \ded{\Gamma^0_{ij}} &=& \Big(
-\,\frac{a^{\prime}}{a}\,\phi^{(2)} \,-\,\frac{1}{2}\,{\psi^{(2)}}^{\prime}\,
-\,\frac{a^{\prime}}{a}\,\psi^{(2)} \,+\, 4\,\Aa \,\left( \phi^{(1)} \right)^2 \,+\,
2\,\phi^{(1)}\,{\psi^{(1)}}^{\prime} + 4\,\Aa\,\phi^{(1)}\,\psi^{(1)} \,+\,
\partial^k \omega^{(1)}\,\partial_k \psi^{(1)} \\
&-&\, \Aa \partial^k \omega^{(1)}\,\partial_k \omega^{(1)} \Big)\,\delta_{ij} \,-\,
\frac{1}{2}\,\partial_i\,\partial_j \omega^{(2)}\,+\, \frac{1}{4}\,\left(
D_{ij} {\chi^{(2)}}^{\prime} + \partial_j\,{\chi_i^{(2)}}^{\prime} +
\partial_i\,{\chi^{(2)}_j}^{\prime} + \left( {\chi^{(2)}_{ij}} \right)^{\prime} \right)
\nonumber  \\
&+& \,\frac{1}{2}\,\frac{a^{\prime}}{a}\,\left( D_{ij}\chi^{(2)} +
\partial_i\,\chi^{(2)}_j +
\partial_j\,\chi_i^{(2)} + \chi^{(2)}_{ij}\,\right)
 \,-\, \frac{1}{4}\,\left( \partial_i \omega^{(2)}_j + \partial_j \omega_i^{(2)}\right)\nonumber \\
&+&\, 2\,\phi^{(1)}\,\partial_i \partial_j \omega^{(1)}\,-\, \partial_i \psi^{(1)}
 \,\partial_j \omega^{(1)} \,-\, \partial_j \psi^{(1)}\,\partial_i \omega^{(1)} \,-\,
 \phi^{(1)}\,D_{ij} {\chi^{(1)}}^{\prime} \,+\, \frac{1}{2}\,\partial^k \omega^{(1)}\,
\partial_i
 D_{kj}\chi^{(1)} \nonumber   \\
&+&\, \frac{1}{2}\,\partial^k \omega^{(1)}\,\partial_j D_{ik}\chi^{(1)}\,-\,
\frac{1}{2}\,\partial^k \omega^{(1)} \,\partial_k D_{ij}\chi^{(1)} \nonumber \, , \\
\ded{\Gamma^i_{0j}} \,&=& -\,\frac{1}{2}\,{\psi^{(2)}}^{\prime}\delta^i_{~j}
\,+\, \frac{1}{4}\,\left(
 D^i_{~j} {\chi^{(2)}}^{\prime} + \partial_j\left(\chi^{i(2)} \right)^{\prime} +
\partial^i\left( \chi^{(2)}_j \right)^{\prime} + \left( {\chi^{i(2)}_{~j}} \right)^{\prime} \right)
\,+ \,\frac{1}{4}\,\left( \partial_j \omega^{i(2)} - \partial^i \omega^{(2)}_j \right) \\
&-& 2\,\psi^{(1)}\,{\psi^{(1)}}^{\prime}\delta^i_{~j}  \,-\,\partial^i \omega^{(1)}
\,\partial_j \phi^{(1)} \,-\, \Aa \partial^i \omega^{(1)} \,\partial_j \omega^{(1)} \,+\,
\psi^{(1)}\,D^i_{~j} {\chi^{(1)}}^{\prime} \,+\, {\psi^{(1)}}^{\prime}D^i_{~j} 
\chi^{(1)} \nonumber \\
&-&
\frac{1}{2}\,D^{ik}\chi^{(1)}\,D_{kj} {\chi^{(1)}}^{\prime} \nonumber \, ,\\
\ded{\Gamma^i_{jk}} \,&=&\,
\frac{1}{2}\,\left(-\partial_j\,\psi^{(2)}\,\delta^i_{~k}
-\partial_k\,\psi^{(2)}\,\delta^i_{~j} + \partial^i\,\psi^{(2)}\,\delta_{jk}\right)
 \,+\, \frac{1}{4} \left( \partial_j D^i_{~k} \chi^{(2)} + \partial_k D^i_{~j} \chi^{(2)}
 - \partial^i D_{jk} \chi^{(2)} \right) \\
&+& \frac{1}{2}\,\partial_j \partial_k\,\chi^{i(2)}
 + \frac{1}{4}\,\left(\partial_j\,\chi^{i(2)}_{~k} + \partial_k\,\chi^{i(2)}_{~j} -
\partial^i\,\chi^{(2)}_{jk}\right)
\,-\, \frac{1}{2}\,\Aa \left( \partial^i \omega^{(2)} + \omega^{i(2)}\right)\,\delta_{jk} \nonumber  \\
&+&\, 2\,\psi^{(1)}\,\left(-\partial_j\,\psi^{(1)}\,\delta^i_{~k}
-\partial_k\,\psi^{(1)}\,\delta^i_{~j} +
\partial^i\,\psi^{(1)}\,\delta_{jk}\right)
\,+\, 2\,\Aa \phi^{(1)}\,\partial^i \omega^{(1)} \,\delta_{jk} \,+\,\partial^i
\omega^{(1)}\,\partial_j \partial_k \omega^{(1)} \nonumber \\
&+&\,{\psi^{(1)}}^{\prime}\partial^i \omega^{(1)} \,\delta_{jk} \,+\,\psi^{(1)}\,\left(
\partial_j D^i_{~k} \chi^{(1)} + \partial_k D^i_{~j} \chi^{(1)} - \partial^i D_{jk}\chi^{(1)} \right)
\,+\, \partial_j \psi^{(1)}\,D^i_{~k} \chi^{(1)}  \nonumber \\
&+&\, \partial_k \psi^{(1)}\,D^i_{~j} \chi^{(1)}\,-\, \partial_m
\psi^{(1)}\,D^{im}\chi^{(1)}\,\delta_{jk} \,-\, \Aa \partial^i \omega^{(1)}\,D_{jk}\chi^{(1)} \,+\,
\Aa \partial^m \omega^{(1)}\,D^i_{~m} \chi^{(1)}\,\delta_{jk} \nonumber   \\
&-&\, \frac{1}{2}\,\partial^i \omega^{(1)} \,D_{jk}{\chi^{(1)}}^{\prime}\,-\,
\frac{1}{2}\,D^{im}\chi^{(1)}\,\partial_j D_{mk}\chi^{(1)} \,-\,
\frac{1}{2}\,D^{im}\chi^{(1)}\,\partial_k D_{mj}\chi^{(1)} 
+ \frac{1}{2}\,D^{im}\chi^{(1)}\,\partial_m D_{jk}\chi^{(1)} \nonumber \,.
\eea
\subsection{The Ricci tensor components}
In a spatially flat Robertson-Walker background the components of the Ricci tensor $R_{\mu\nu}$ are given by
\begin{equation}
R_{00}\,=\,-\,3\,\Ac \,+\,3\,\Ab \,; \qquad R_{0i}\,=\,0\,;
\end{equation}
\begin{equation}
R_{ij}\,=\,\left[ \Ac \,+\,\Ab \right]\,\delta_{ij}\,.
\end{equation}
The first-order perturbed Ricci tensor components are
\bea
\deu {R_{00}}&=& \frac{a^{\prime}}{a}\partial_i\partial^i \omega^{(1)} +
\partial_i\partial^i {\omega^{(1)}}^{\prime} + \partial_i\partial^i \phi^{(1)} +
3{\psi^{(1)}}^{\prime\prime} + 3\frac{a^{\prime}}{a}{\psi^{(1)}}^{\prime} +
3\frac{a^{\prime}}{a}{\phi^{(1)}}^{\prime} \, ,\\
\deu {R_{0i}}&=& \frac{a^{\prime\prime}}{a}\partial_i \omega^{(1)} +
\left(\frac{a^{\prime}}{a}\right)^2\partial_i \omega^{(1)} +
2\partial_i{\psi^{(1)}}^{\prime} + 2\frac{a^{\prime}}{a}\partial_i \phi^{(1)} +
\frac{1}{2}\partial_k D^k_{~i} {\chi^{(1)}}^{\prime} \, , \\
\deu {R_{ij}}&=& \left[  -\frac{a^{\prime}}{a}{\phi^{(1)}}^{\prime} -
5\frac{a^{\prime}}{a}{\psi^{(1)}}^{\prime} - 2\frac{a^{\prime\prime}}{a}\phi^{(1)}
-2\left(\frac{a^{\prime}}{a}\right)^2\phi^{(1)} 
-2\frac{a^{\prime\prime}}{a}\psi^{(1)} -
2\left(\frac{a^{\prime}}{a}\right)^2\psi^{(1)} - {\psi^{(1)}}^{\prime\prime} +
\partial_k\partial^k\psi^{(1)} \right. \\
&-& \left.  \frac{a^{\prime}}{a}\partial_k\partial^k \omega^{(1)} \right] \delta_{ij}
-\partial_i\partial_j {\omega^{(1)}}^{\prime} +
\frac{a^{\prime}}{a}D_{ij}{\chi^{(1)}}^{\prime} +
\frac{a^{\prime\prime}}{a}D_{ij}\chi^{(1)} +
\left(\frac{a^{\prime}}{a}\right)^2 D_{ij}\chi^{(1)} 
+ \frac{1}{2}D_{ij}{\chi^{(1)}}^{\prime\prime} + \partial_i\partial_j \psi^{(1)}\nonumber \\
&-& \partial_i\partial_j \phi^{(1)} - 2\frac{a^{\prime}}{a}\partial_i\partial_j \omega^{(1)} 
+ \frac{1}{2}\partial_k\partial_iD^k_{~j} \chi^{(1)} +
\frac{1}{2}\partial_k\partial_j D^k_{~i} \chi^{(1)} -
\frac{1}{2}\partial_k\partial^k D_{ij} \chi^{(1)} \nonumber \, .
\eea
At second order we obtain
\bea
\ded R_{00}&=& +\, \frac{3}{2} \Aa\ \phi^{(2)} \,+\, \frac{1}{2}\,\La\
\phi^{(2)} \,+\, \frac{3}{2} \Aa\ {\psi^{(2)}}^{\prime}\,+\,
\frac{3}{2}\,{\psi^{(2)}}^{\prime\prime}\,+\,\frac{1}{2}\,\Aa\
\partial_k \,\partial^k \omega^{(2)} +  \frac{1}{2}\, \partial_k \,\partial^k {\omega^{(2)}}^{\prime} \, \\
&-&\, 6
\,\Aa\ \phi^{(1)}\,{\phi^{(1)}}^{\prime} \,-\,
\partial^k \phi^{(1)} \,\partial_k \phi^{(1)} \,-\, 3
{\phi^{(1)}}^{\prime} {\psi^{(1)}}^{\prime} \,+\, 2\,\psi^{(1)} \,\La\ \phi^{(1)} -\, \partial_k \psi^{(1)} \,
\partial^k \phi^{(1)}\, \nonumber \\
&+& 6 \,\Aa\
\psi^{(1)}\,{\psi^{(1)}}^{\prime}\,+\, 6\psi^{(1)}\,{\psi^{(1)}}^{\prime\prime} \,+\, 3
\left( {\psi^{(1)}}^{\prime} \right)^2\,-\,{\phi^{(1)}}^{\prime}\La\ \omega^{(1)} + \Aa\ \partial^k 
\omega^{(1)} \,\partial_k \phi^{(1)}\,  \nonumber \\ 
&+& \,\Ac\ \partial^k \omega^{(1)}
\partial_k \omega^{(1)} + \Ab \partial^k \omega^{(1)} \,\partial_k \omega^{(1)}\,- \, \Aa \partial_k
\psi^{(1)}\,
\partial^k \omega^{(1)} +\, 2 \,\Aa \psi^{(1)} \,\La\omega^{(1)} \, \nonumber \\
&-&\, \partial_k \psi^{(1)} \,\partial^k {\omega^{(1)}}^{\prime}
 +\, 2 \,\psi^{(1)}\, \La {\omega^{(1)}}^{\prime} \,+\, 3 \,\Aa\ \partial^k \omega^{(1)} \,
\partial_k {\omega^{(1)}}^{\prime} -\partial_k \phi^{(1)} \,\partial_i D^{ik} \chi^{(1)}\, \nonumber \\
&-&\, \partial_i
\partial_k \phi^{(1)} \, D^{ik} \chi^{(1)} \,- \Aa\
\partial_i\partial_k \omega^{(1)}\, D^{ik} \chi^{(1)} \,-\, \Aa \partial_k \omega^{(1)}
\,\partial_i D^{ik} \chi^{(1)} -\, \partial_k {\omega^{(1)}}^{\prime} \,\partial_i D^{ik} \chi^{(1)} \nonumber \\
&-& \partial_i
\partial_k {\omega^{(1)}}^{\prime} \,D^{ik} \chi^{(1)}   \,+\, \frac{1}{2} D^{ik} \chi^{(1)} \,
D_{ki} {\chi^{(1)}}^{\prime\prime}
+\, \frac{1}{4} D^{ik} {\chi^{(1)}}^{\prime}\,D_{ki} {\chi^{(1)}}^{\prime}
\,+\, \frac{1}{2} \Aa\ D^{ik} \chi^{(1)} \,D_{ki} {\chi^{(1)}}^{\prime} \,. \nonumber \\
\ded R_{0i} &=& +\,\Aa \partial_i \phi^{(2)} \,+\, \partial_i
{\psi^{(2)}}^{\prime} \,+\, \frac{1}{4} \partial_k \,D^k_{~i} {\chi^{(2)}}^{\prime} \,+\,
\frac{1}{4}
\La \,{\chi^{(2)}_i}^{\prime} \,-\,\frac{1}{4} \La \,\omega^{(2)}_i \\ 
&+&\frac{1}{2} \Ac (\,\partial_i \omega^{(2)} + \omega^{(2)}_i\,)
+\, \frac{1}{2}
\Ab (\,\partial_i \omega^{(2)} + \omega^{(2)}_i\,) \,-\, 4\,\Aa\, \phi^{(1)}\, \partial_i \phi^{(1)} 
- 2 {\psi^{(1)}}^{\prime} \partial_i \phi^{(1)} \nonumber \\
&+&\,4
\,{\psi^{(1)}}^{\prime}\,\partial_i \psi^{(1)} \,+\, 4 \,\psi^{(1)} \,\partial_i
{\psi^{(1)}}^{\prime}\,-\,2\, \Ac \phi^{(1)} \partial_i \omega^{(1)} \,-\, 2\,\Ab
\phi^{(1)} \,\partial_i \omega^{(1)} - \Aa {\phi^{(1)}}^{\prime} \partial_i \omega^{(1)} \nonumber \\
&-&\, \La \omega^{(1)}\,\partial_i \phi^{(1)}
\,-\,
\partial^k \omega^{(1)} \,\partial_i\partial_k \phi^{(1)} \,+\, \partial_k \phi^{(1)} \,
\partial_i\partial_k \omega^{(1)}  \,-\, \partial^k \omega^{(1)} \,\partial_i\partial_k 
{\omega^{(1)}}^{\prime}- \Aa \La \omega^{(1)}
\partial_i \omega^{(1)} \nonumber \\
&-& {\psi^{(1)}}^{\prime\prime}\partial_i \omega^{(1)} \,-\,5 \,\Aa
{\psi^{(1)}}^{\prime}\partial_i \omega^{(1)} \,-\, \frac{1}{2} \partial^k \phi^{(1)} \,D^{ik}
{\chi^{(1)}}^{\prime}\,+\, \psi^{(1)} \,\partial_k\, D^k_{~i} {\chi^{(1)}}^{\prime}
+ {\psi^{(1)}}^{\prime} \,\partial_k \,D^k_{~i} \chi^{(1)} \nonumber \\
&-&\, \frac{1}{2}
\partial_k \psi^{(1)} \,D^k_i {\chi^{(1)}}^{\prime} \,+\,
\partial_k {\psi^{(1)}}^{\prime} \,D^k_{~i} \chi^{(1)}
\,+\, \Aa \partial^k \omega^{(1)} \,D_{ik}{\chi^{(1)}}^{\prime}  
+ \frac{1}{2} \partial^k \omega^{(1)} \,D_{ik} {\chi^{(1)}}^{\prime\prime} \nonumber \\
&-& \frac{1}{2} \partial_k
D^{km} \chi^{(1)} \,D_{mi} {\chi^{(1)}}^{\prime}
 \,-\, \frac{1}{2} D^{km} \chi^{(1)} \,\partial_k D_{mi} {\chi^{(1)}}^{\prime} + 
\frac{1}{2} D^{km} {\chi^{(1)}}^{\prime} \,\partial_i D_{mk} \chi^{(1)} +
\frac{1}{4} D^{km} \chi^{(1)} \,\partial_i D_{mk} {\chi^{(1)}}^{\prime} \nonumber \, . 
\eea
The purely spatial part of $\ded \R_{\mu\nu}$ is very long, and for 
simplicity has been divided 
into two parts, a diagonal part $\ded {R^d_{ij}}$ which is proportional to $\delta_{ij}$, 
and a non-diagonal part ${R^{nd}_{ij}}$. 
\bea
\ded {R^d_{ij}} &=& \Bigg[ -\,
 \Ab \phi^{(2)} \,-\,\frac{1}{2} \Aa {\phi^{(2)}}^{\prime} \,-\, \Ac \phi^{(2)}\,-\,
  \,-\, \frac{5}{2}\,\Aa {\psi^{(2)}}^{\prime}
\,-\, \Ab \psi^{(2)} \,-\,\frac{1}{2} {\psi^{(2)}}^{\prime\prime} \\
&-& \Ac \psi^{(2)} \,+\, \frac{1}{2}\, \La \psi^{(2)} \,-\,\frac{1}{2} \Aa \La \omega^{(2)}
\,+\,4 \left(\Ab + \Ac\right) \left( \phi^{(1)} \right)^2 \,+\,  4\,
\Aa \phi^{(1)}\,{\phi^{(1)}}^{\prime} \nonumber \\
&+&10\,\Aa \phi^{(1)}\,{\psi^{(1)}}^{\prime} \,+\,2\,\Aa {\phi^{(1)}}^{\prime}\psi^{(1)} \,+ \,
{\phi^{(1)}}^{\prime}{\psi^{(1)}}^{\prime}
\,+\,2\,\phi^{(1)}\,{\psi^{(1)}}^{\prime\prime} \,
+\, 4 \left(\Ab + \Ac\right)\phi^{(1)}\,\psi^{(1)} \nonumber \\
&+&\partial_k \psi^{(1)}\,\partial^k \phi^{(1)} \, +\, \left( {\psi^{(1)}}^{\prime} \right)^2 \,+\,
\partial_k \psi^{(1)} \, \partial^k \psi^{(1)} \,+\,
2\,\psi^{(1)} \La \psi^{(1)} \,+\,\Aa \partial_k \phi^{(1)}\,\partial^k \omega^{(1)} \nonumber \\
&+& 2\,\Aa \phi^{(1)}\,\La \omega^{(1)} \, - \Ab
\partial_k \omega^{(1)} \,\partial^k \omega^{(1)} \,-\, \Ac \partial_k \omega^{(1)}\,\partial^k \omega^{(1)}
\,-\, \Aa \partial_k \omega^{(1)}\,\partial^k {\omega^{(1)}}^{\prime}\nonumber \\
&+& 3\,\Aa
\partial_k \omega^{(1)}\,\partial^k \psi^{(1)} \,+\, 2\,\partial_k
{\psi^{(1)}}^{\prime}\,\partial^k \omega^{(1)} \,+\, {\psi^{(1)}}^{\prime}\La \omega^{(1)} \,+\,
\partial_k \psi^{(1)}\,\partial^k {\omega^{(1)}}^{\prime} 
- \partial_m \psi^{(1)} \,\partial_k D^{km} \chi^{(1)}\nonumber \\
&-& \partial_k
\partial_m \psi^{(1)} \,D^{km} \chi^{(1)}
\,+\, \Aa \partial_m \partial^k \omega^{(1)}\,D^m_k \chi^{(1)} 
+ \Aa \partial^k \omega^{(1)} \,\partial_m D^m_k \chi^{(1)} - \frac{1}{2} \Aa
D^{mk} \chi^{(1)} \, D_{km}
{\chi^{(1)}}^{\prime} \Bigg]\,\delta_{ij} \,,\nonumber \\
\ded {R^{nd}_{ij}} &=& 
-\frac{1}{2} \,\partial_i\partial_j \phi^{(2)}
\,+\,\frac{1}{2}\,\partial_i\partial_j \psi^{(2)} \,-\, \Aa
\partial_i\partial_j \omega^{(2)} \,-\, \frac{1}{2}\,\partial_i\partial_j
{\omega^{(2)}}^{\prime} \,-\, \frac{1}{2} \Aa \left( \partial_i \,\omega^{(2)}_j +
\partial_j\, \omega^{(2)}_i\right) \\
&-& \frac{1}{4} \left( \partial_i \,{\omega^{(2)}_j}^{\prime} + \partial_j\,
{\omega^{(2)}_i}^{\prime} \right) 
+\,\frac{1}{2} \left(\Ab + \Ac\right)
\left( D_{ij} \chi^{(2)} +
\partial_i \chi^{(2)}_j
+ \partial_j \chi^{(2)}_i + \chi^{(2)}_{ij}\right) \nonumber \\
&+& \frac{1}{2} \Aa \left( D_{ij} {\chi^{(2)}}^{\prime} +
\partial_i {\chi^{(2)}_j}^{\prime}
+ \partial_j {\chi^{(2)}_i}^{\prime} +\left( {\chi^{(2)}_{ij}} \right)^{\prime}\right) \,+\,
\frac{1}{2}
\partial_k\partial_i\,D^k_{~j} \chi^{(2)} \,-\,\frac{1}{4} \La D_{ij} \chi^{(2)}\nonumber \\
&-& \frac{1}{4} \La \chi^{(2)}_{ij} \,+\, \frac{1}{4}\left( D_{ij}
{\chi^{(2)}}^{\prime\prime} +
\partial_i {\chi^{(2)}_j}^{\prime\prime}
+ \partial_j {\chi^{(2)}_i}^{\prime\prime} +
\left({\chi^{(2)}_{ij}}\right)^{\prime\prime}\right) \,
+ \, \partial_i \phi^{(1)}\, \partial_j \phi^{(1)}\nonumber \\
&+& 2\,\phi^{(1)}\,\partial_i\partial_j \phi^{(1)} \,-\,\partial_j \phi^{(1)}\,
\partial_i\psi^{(1)} \,-\, \partial_i \phi^{(1)}\,\partial_j \psi^{(1)} \,+\,
3\,\partial_i \psi^{(1)}\,\partial_j \psi^{(1)} \,+\, 2 \,\psi^{(1)}\,\partial_i\partial_j \psi^{(1)} 
\nonumber \\
&+& 4 \Aa \phi^{(1)}\,\partial_i\partial_j \omega^{(1)} \,+\,
{\phi^{(1)}}^{\prime}\,\partial_i\partial_j \omega^{(1)} \,+\, 2\,\phi^{(1)}
\,\partial_i\partial_j {\omega^{(1)}}^{\prime} \,+\, \La \omega^{(1)}
\,\partial_i\partial_j \omega^{(1)} \nonumber \\
&-& \partial_j
\partial^k \omega^{(1)}\,\partial_i \partial_k \omega^{(1)}
\,-\,2 \, \Aa \partial_i \psi^{(1)} \,\partial_j \omega^{(1)} \,-\, 2\,\Aa
\partial_i \omega^{(1)}\,\partial_j \psi^{(1)} \,-\, \partial_i
{\psi^{(1)}}^{\prime}\,\partial_j \omega^{(1)} \nonumber \\
&-& \partial_j {\psi^{(1)}}^{\prime}\,\partial_i \omega^{(1)} \,-\, \partial_i \psi^{(1)}
\,\partial_j {\omega^{(1)}}^{\prime} \,-\, \partial_j \psi^{(1)} \,\partial_i
{\omega^{(1)}}^{\prime} \,+\, {\psi^{(1)}}^{\prime}\,\partial_i\partial_j \omega^{(1)} \,-\, 2
\,\Ab \phi^{(1)}\,D_{ij} \chi^{(1)} \nonumber \\
&-& 2 \,\Ac \phi^{(1)}\,D_{ij} \chi^{(1)} \,-\, 2\,\Aa \phi^{(1)}\,D_{ij} {\chi^{(1)}}^{\prime} \,-\,
\Aa {\phi^{(1)}}^{\prime}\, D_{ij} \chi^{(1)} \, -\,\frac{1}{2} {\phi^{(1)}}^{\prime}\,D_{ij}
{\chi^{(1)}}^{\prime} \,-\, \phi^{(1)}\,D_{ij} {\chi^{(1)}}^{\prime\prime}\nonumber \\
&+& \frac{1}{2}
\partial_k \phi^{(1)} \, \partial_i D^k_{~j} \chi^{(1)} \,+\, \frac{1}{2} \partial_k \phi^{(1)}
\, \partial_j D^k_{~i} \chi^{(1)} \,-\, \frac{1}{2} \partial_k \phi^{(1)} \,
\partial^k D_{ij} \chi^{(1)}  \,-\, 3\,\Aa
{\psi^{(1)}}^{\prime}\,D_{ij} \chi^{(1)} \nonumber \\
&+& \frac{1}{2} {\psi^{(1)}}^{\prime}\, D_{ij} {\chi^{(1)}}^{\prime}\,+\,
\frac{1}{2}
\partial_k \psi^{(1)}\,
\partial_i D^k_{~j} \chi^{(1)} \,+\, \frac{1}{2} \partial_k \psi^{(1)}\,
\partial_j D^k_{~i} \chi^{(1)} \,-\,\frac{3}{2}\, \partial_k \psi^{(1)} \,\partial^k D_{ij}
\chi^{(1)}\nonumber \\
&+& \psi^{(1)}\,\partial_k\partial_i D^k_{~j} \chi^{(1)} \,+\,
\psi^{(1)}\,\partial_k\partial_j D^k_{~i} \chi^{(1)} \,-\,
\psi^{(1)}\,\partial_k\partial^k D_{ij} \chi^{(1)} +\,
\partial_i\psi^{(1)}\,\partial_k D^k_{~j} \chi^{(1)}  \nonumber \\
&+&
\partial_j\psi^{(1)}\,\partial_k D^k_{~i} \chi^{(1)}
\, +\, \partial_k\partial_i \psi^{(1)}\,D^k_{~j} \chi^{(1)} \,+\,
\partial_k\partial_j \psi^{(1)}\,D^k_{~i} \chi^{(1)}
\, +\, \frac{1}{2}\,\partial_k\partial_i \omega^{(1)} \, D^k_{~j} {\chi^{(1)}}^{\prime}
\nonumber \\
&+& \frac{1}{2}\,\partial_k\partial_j \omega^{(1)} \, D^k_{~i} {\chi^{(1)}}^{\prime}
\,-\, \frac{1}{2} \,\partial_k\partial^k \omega^{(1)}\, D_{ij} {\chi^{(1)}}^{\prime}
\,+\,\frac{1}{2}\, \partial^k \omega^{(1)}\, \partial_i D_{kj} {\chi^{(1)}}^{\prime}
+\,\frac{1}{2}\, \partial^k \omega^{(1)}\, \partial_j D_{ki}
{\chi^{(1)}}^{\prime}\nonumber \\
&-&\partial^k \omega^{(1)} \,\partial_k D_{ij} {\chi^{(1)}}^{\prime}\,+\, \frac{1}{2}
\,\partial^k {\omega^{(1)}}^{\prime} \,\partial_i D_{kj} \chi^{(1)} \,+\, \frac{1}{2}
\,\partial^k {\omega^{(1)}}^{\prime} \,\partial_j D_{ki} \chi^{(1)} \,-\, \frac{1}{2}
\,\partial^k {\omega^{(1)}}^{\prime} \,\partial_k D_{ij} \chi^{(1)}\nonumber \\
&+& \Aa \partial^k \omega^{(1)} \,\partial_i D_{kj} \chi^{(1)} \,+\, \Aa \partial^k
\omega^{(1)} \,\partial_j D_{ki} \chi^{(1)} \,-\, \Aa
\partial^k \omega^{(1)} \,\partial_k D_{ij} \chi^{(1)} \,-\, \Aa
\partial_k\partial^k \omega^{(1)} \,D_{ij} \chi^{(1)} \nonumber \\
&-&\frac{1}{2}\, D^k_i {\chi^{(1)}}^{\prime} \, D_{kj} {\chi^{(1)}}^{\prime} \,-\,
\frac{1}{2} \partial_i D_{mj} \chi^{(1)} \,
\partial_k D^{km} \chi^{(1)} \,-\, \frac{1}{2} \partial_j D_{mi} \chi^{(1)} \,
\partial_k D^{km} \chi^{(1)}  \nonumber \\
&+& \frac{1}{2} \partial_m D_{ij} \chi^{(1)} \, \partial_k D^{km} \chi^{(1)}
-\,\frac{1}{2} \partial_k\partial_i D_{mj} \chi^{(1)} \, D^{km} \chi^{(1)} \,-\,
\frac{1}{2} \partial_k\partial_j D_{mi} \chi^{(1)} \, D^{km} \chi^{(1)} \nonumber \\
&+&\frac{1}{2} \partial_k\partial_m D_{ij} \chi^{(1)} \, D^{km} \chi^{(1)} \,+\,
\frac{1}{2} D^{km} \chi^{(1)} \,\partial_i\partial_j D_{km} \chi^{(1)} \,+\,
\frac{1}{4} \partial_i D^{mk} \chi^{(1)} \,\partial_j D_{mk} \chi^{(1)} \nonumber \,. 
\eea
\subsection{Ricci scalar} 
At zeroth order the Ricci scalar $R$ is given by
\be
R=\frac{6}{a^2} \frac{a^{\prime\prime}}{a} \, .
\ee
The first-order perturbation of $R$ is
\bea
\deu R &=& \frac{1}{a^2} \left( 
-6\frac{a^{\prime}}{a}\partial_i\partial^i \omega^{(1)} -
2\partial_i\partial^i {\omega^{(1)}}^{\prime} - 2\partial_i\partial^i \phi^{(1)}
-6{\psi^{(1)}}^{\prime\prime}
 - 6\frac{a^{\prime}}{a}{\phi^{(1)}}^{\prime}
-18\frac{a^{\prime}}{a}{\psi^{(1)}}^{\prime} \right. \\
&-& \left. 12\frac{a^{\prime\prime}}{a}\phi^{(1)} 
+  4\partial_i\partial^i\psi^{(1)} +
\partial_k\partial^i D^k_{~i} \chi^{(1)} \right) \, . \nonumber 
\eea
At second order we find
\bea
\ded R &=& -\, \La \phi^{(2)} \,-\, 3\,\Aa {\phi^{(2)}}^{\prime} \,-\, 6\,\Ac \phi^{(2)}
\,+\,
2\,\La \psi^{(2)} \,-\, 9\,\Aa {\psi^{(2)}}^{\prime} \,-\, 3\,{\psi^{(2)}}^{\prime\prime}
\,-\, \La {\omega^{(2)}}^{\prime} \\
&-& 3\,\Aa \La \omega^{(2)}  \,+\, \frac{1}{2}\,\partial_k\partial_i
\,D^{ki}\chi^{(2)} \,+\,24\,\Ac \left( \phi^{(1)} \right)^2 \,+\,
2\,\partial_k \phi^{(1)}\,\partial^k \phi^{(1)} \,+\, 4\,\phi^{(1)}\,\La \phi^{(1)}  \nonumber \\
&+& 24\,\Aa \phi^{(1)}\,{\phi^{(1)}}^{\prime}\,+\, 6\,{\phi^{(1)}}^{\prime}{\psi^{(1)}}^{\prime}
\,+\,36\,\Aa \phi^{(1)}\,{\psi^{(1)}}^{\prime}\,+\, 2\,
\partial_k \psi^{(1)} \,\partial^k \phi^{(1)} \,-\, 4\,\psi^{(1)} \,\La \phi^{(1)} \nonumber \\
&+& 12
\,\phi^{(1)}\,{\psi^{(1)}}^{\prime\prime}\,-\,12\,\psi^{(1)}\,{\psi^{(1)}}^{\prime\prime}
\,-\,36\,\Aa {\psi^{(1)}}^{\prime}\psi^{(1)} \,+\, 6\,\partial_k
\psi^{(1)}\,\partial^k \psi^{(1)} \,+\, 16\,\psi^{(1)}\,\La \psi^{(1)} \nonumber \\
&+& 6\,\Aa
\partial^k \omega^{(1)}\,\partial_k \phi^{(1)} \,+\, 12\,\Aa \phi^{(1)}\,\La \omega^{(1)} \,
+\, 4\,\phi^{(1)}\,\La
{\omega^{(1)}}^{\prime} \,+\, 2\,{\phi^{(1)}}^{\prime}\La \omega^{(1)} \,\nonumber \\
&-& 5\,\Ac \partial_k \omega^{(1)}\,\partial^k \omega^{(1)} \,-\, 6\,\Aa \partial_k
\omega^{(1)}\,\partial^k {\omega^{(1)}}^{\prime} \,+\, \La \omega^{(1)}\,\La \omega^{(1)} \,-\,
\partial^i\partial^k \omega^{(1)}\,\partial_i
\partial_k \omega^{(1)} \nonumber \\
&+& 8\,\partial_k \omega^{(1)}\,\partial^k {\psi^{(1)}}^{\prime}\,+\,2\,\partial_k
{\omega^{(1)}}^{\prime}\partial^k \psi^{(1)} \,-\,
4\,\psi^{(1)}\,\La {\omega^{(1)}}^{\prime} \,-\, 12\,\Aa \psi^{(1)}\,\La \omega^{(1)} \nonumber \\
&+& 4\,{\psi^{(1)}}^{\prime}\La \omega^{(1)} \,+\, 2\,\partial_k \phi^{(1)}\,\partial_i
D^{ik} \chi^{(1)} \,+\, 2\, \partial_i
\partial_k \phi^{(1)} \,D^{ik} \chi^{(1)} \,
+\, 4\,\psi^{(1)}\,\partial_k\partial_i\,D^{ki} \chi^{(1)} \nonumber \\
&-& 2\,
\partial_k
\partial_i \psi^{(1)}\,D^{ik}\chi^{(1)} \,
+\, 3\, \partial_k \omega^{(1)}\,\partial^i D^k_{~i} {\chi^{(1)}}^{\prime}
 \,+\, 6\,\Aa
\partial^k \omega^{(1)}\,\partial_i D^i_{~k} \chi^{(1)}\,
+\,2\,\partial_i {\omega^{(1)}}^{\prime}\partial_k D^{ik}\chi^{(1)}
 \nonumber \\
&+& 2\, \partial_k\partial_i {\omega^{(1)}}^{\prime}D^{ik}\chi^{(1)}  \,+\, 6\,\Aa
 \partial_k\partial_i \omega^{(1)}\,D^{ki}\chi^{(1)} \,-\,
 D^{ik}\chi^{(1)}\,D_{ik}{\chi^{(1)}}^{\prime\prime} \,-\,
 \frac{3}{4}\,D^{ik}{\chi^{(1)}}^{\prime}D_{ki}{\chi^{(1)}}^{\prime} \nonumber \\
&-& 3\,\Aa
 D^{ik}\chi^{(1)}\,D_{ik}{\chi^{(1)}}^{\prime} \,
-\, 2 \,\partial_k\partial^i\,D_{mi}\chi^{(1)}\,D^{km}\chi^{(1)} \, +\, \La
 D_{im}\chi^{(1)}\,D^{mi}\chi^{(1)} \nonumber \\
&-&\, \partial_k D^{km}\chi^{(1)}\,\partial^i D_{mi}\chi^{(1)}
 \,+\, \frac{1}{4}\,\partial^i D^{km}\chi^{(1)}\,\partial_i D_{mk}\chi^{(1)} \, . \nonumber 
\eea

\subsection{The Einstein tensor components}
The Einstein tensor in a spatially flat Robertson-Walker background is given by
\bea
G^0_{~0} &=&  -\frac{3}{a^2} \left( \frac{a^{\prime}}{a} \right)^2 \, ,\\
G^i_{~j} &=& -\frac{1}{a^2} \left( 2 \frac{a^{\prime\prime}}{a}
-\Ab \right)~\delta^i_{~j} \, ,\\
G^0_{~i} &=& G^i_{~0} = 0 \, .
\eea
The first-order perturbations of the Einstein tensor components are
\bea
\deu {G_{~0}^{0}}&=& \frac{1}{a^2}\Bigg[ 6\,\Ab \phi^{(1)} \,+\,6\,\Aa{\psi^{(1)}}^{\prime}\,+\, 2\,\Aa \La
 \omega^{(1)}\,-\, 2\,\La \psi^{(1)} \,-\, \frac{1}{2}\,\partial_k \partial^i
 \,D^k_i \chi^{(1)} \Bigg] \, , \\
\deu {G^0_{~i}}&=& \frac{1}{a^2}\left( -\,2\,\Aa \partial_i \phi^{(1)} \,-\, 2\,\partial_i
{\psi^{(1)}}^{\prime} \,-\, \frac{1}{2}\,\partial_k D^k_{~i}{\chi^{(1)}}^{\prime} \right)\, , \\
\deu {G^i_{~j}} &=& \frac{1}{a^2} \Bigg[ \left( 2\,\Aa {\phi^{(1)}}^{\prime} \,+\, 4\,\Ac \phi^{(1)} \,-\,
2\,\Ab \phi^{(1)} \,+\, \La \phi^{(1)} \,+\, 4\,\Aa {\psi^{(1)}}^{\prime} \,+\, 
2\,{\psi^{(1)}}^{\prime\prime} 
\right. \\
&-& \left. \La \psi^{(1)} \,+\, 2\,\Aa \La \omega^{(1)} \,+\, \La {\omega^{(1)}}^{\prime} \,+\,
\frac{1}{2}\partial_k \partial^m D^k_{~m}
\chi^{(1)} \right)\delta_{~j}^{i} \nonumber \\
&-& \partial^i\partial_j \phi^{(1)} \,+\, \partial^i\partial_j \psi^{(1)} \,-\,
2\,\Aa \partial^i\partial_j \omega^{(1)} \,-\, \partial^i\partial_j
{\omega^{(1)}}^{\prime} \,+\, \Aa D^i_{~j} {\chi^{(1)}}^{\prime} \,+\, \frac{1}{2}\,D^i_{~j}
{\chi^{(1)}}^{\prime\prime} \nonumber \\
&+& \frac{1}{2}\,\partial_k\partial^i \,D^k_{~j} \chi^{(1)} \,+\,
\frac{1}{2}\,\partial_k\partial_j \,D^{ik} \chi^{(1)} \,-\,
\frac{1}{2}\,\partial_k\partial^k \,D^i_{~j} \chi^{(1)}\Bigg] \,. \nonumber 
\eea  
The second-order perturbed Einstein tensor components are given by 
\bea
\ded {G_{~0}^0} &=& \frac{1}{a^2} \Big( 3\Ab \phi^{(2)}\,+\,3\,\Aa {\psi^{(2)}}^{\prime} \,
-\, \La \psi^{(2)}
\,+\,\Aa \La \omega^{(2)} \,-\,\frac{1}{4} \partial_k \partial_i\,D^{ki} \chi^{(2)}
\\
&-&\, 12 \left( \Aa \right)^2 \left( \phi^{(1)} \right)^2 - 12\,\Aa\,\phi^{(1)}\,{\psi^{(1)}}^{\prime} 
-\,3\,\partial_i \psi^{(1)}\,\partial^i \psi^{(1)} \,-\, 8\,\psi^{(1)}\ \La \psi^{(1)}
\,+\, 12\,\Aa \psi^{(1)}\,{\psi^{(1)}}^{\prime} \nonumber \\
&-& 3\,\left( {\psi^{(1)}}^{\prime} \right)^2 \,+\, 4\,\Aa\,\phi^{(1)}\,\La \omega^{(1)}\,-\, 2\,\Aa
\partial_k \omega^{(1)}\,\partial^k \phi^{(1)} \,-\,
\frac{1}{2} \Ac\,\partial_k \omega^{(1)}\,\partial^k \omega^{(1)} \nonumber \\
&+& \frac{1}{2}\,\partial_i\partial_k \omega^{(1)}\,\partial^i\partial^k \omega^{(1)}
\,-\, \frac{1}{2}\,\partial_k\partial^k \omega^{(1)}\,\partial_k\partial^k \omega^{(1)}
\,-\,2\,\Aa \partial_k \psi^{(1)}\,\partial^k \omega^{(1)} \,+\, 4\,\Aa \psi^{(1)}\,\La
\omega^{(1)} \nonumber \\
&-& 2\, \partial_k \omega^{(1)} \,\partial^k {\psi^{(1)}}^{\prime} \,-\,
2{\psi^{(1)}}^{\prime} \La \omega^{(1)} \,-\, \phi^{(1)}\,\partial_i\partial^k \,D^i_{~k} \chi^{(1)} \,-\,
2\,\psi^{(1)}
\partial_k\partial^i \,D^k_{~i} \chi^{(1)} \nonumber \\
&+& \partial_k \partial_i \psi^{(1)}\,D^{ki} \chi^{(1)} \,-\,2\,\Aa
\partial_i\partial_k \omega^{(1)}\,D^{ik} \chi^{(1)} \,-\,2\,\Aa \partial_k \omega^{(1)} \,\partial_i\,
D^{ik} \chi^{(1)}
\,-\,\partial_k \omega^{(1)}\,\partial^i D^k_{~i} {\chi^{(1)}}^{\prime} \nonumber \\
&-& \frac{1}{2}\,\La \,D_{mk} \chi^{(1)} \, D^{km}\chi^{(1)} \,+\,
\partial_m\partial^k \,D_{ik} \chi^{(1)}\,D^{im} \chi^{(1)} \,+\,
\frac{1}{2} \,\partial_k D^{km} \chi^{(1)}\, \partial^i D_{mi} \chi^{(1)} \nonumber \\
&-& \frac{1}{8}\,\partial^i D^{km}\chi^{(1)} \, \partial_i D_{km} \chi^{(1)} \,+\,
\frac{1}{8}\,D^{ik} {\chi^{(1)}}^{\prime} \,D_{ki} {\chi^{(1)}}^{\prime} \,+\, \Aa
D^{ki} \chi^{(1)} \, D_{ik} {\chi^{(1)}}^{\prime} \Big)\, , \nonumber \\
\ded {G_{~0}^i} &=&\frac{1}{a^2} \Big( \Aa \partial^i \phi^{(2)} \,+\,\partial^i {\psi^{(2)}}^{\prime}
\,+\, \frac{1}{4}
\partial_k\,D^{ki} {\chi^{(2)}}^{\prime} \,+\, \frac{1}{4} \La {\chi^{i(2)}}^{\prime}
\,-\, \frac{1}{4} \La \omega^{i(2)} \,-\,\Ac \partial^i \omega^{(2)} \\
&-& \Ac \omega^{i(2)} \,+\,2 \Ab \partial^i \omega^{(2)} \,+\,2 \Ab \omega^{i(2)} \,-\, 4\, \Aa
\phi^{(1)}\,\partial^i \phi^{(1)} \,+\, 4\,\Aa\,\psi^{(1)}\,\partial^i \phi^{(1)}
\nonumber \\
&-& 2 \,{\psi^{(1)}}^{\prime}\partial^i \phi^{(1)}\,+\,
4\,{\psi^{(1)}}^{\prime}\partial^i \psi^{(1)} \,+\, 8\,\psi^{(1)}\,\partial^i
 {\psi^{(1)}}^{\prime}\,-\, \partial^i \phi^{(1)}\,\La \omega^{(1)} \,-\,\partial^k \omega^{(1)} \,
\partial^i\partial_k \phi^{(1)} \nonumber \\
&+& \La \phi^{(1)}
 \,\partial^i \omega^{(1)} \,+\, \partial^i\partial_k \omega^{(1)} \,\partial^k \phi^{(1)} \,+\,
 4\,\Ac \phi^{(1)}\,\partial^i \omega^{(1)} \,-\, 8 \Ab \phi^{(1)}\,\partial^i \omega^{(1)} \nonumber \\
&+& 2 \,\Aa {\phi^{(1)}}^{\prime} \partial^i \omega^{(1)} \,+\, \La {\omega^{(1)}}^{\prime}
 \,\partial^i \omega^{(1)} \,-\,\partial^k \omega^{(1)} \,\partial^i\partial_k
 {\omega^{(1)}}^{\prime}  \,+\,
 2\,{\psi^{(1)}}^{\prime \prime}\partial^i \omega^{(1)} \nonumber \\
&+&\,8\,\Ab \psi^{(1)}\,\partial^i \omega^{(1)}
 \,-\,4\,\Ac\,\psi^{(1)}\,\partial^i \omega^{(1)} \, -\, 2\,\Aa
{\psi^{(1)}}^{\prime}\partial^i \omega^{(1)} \,-\,
 \frac{1}{2}\,\partial^k \phi^{(1)} \,D^i_{~k} {\chi^{(1)}}^{\prime} \nonumber \\
&-&2\,\Aa\,\partial_k \phi^{(1)}\,D^{ki}\chi^{(1)}
 \,-\,
 \frac{1}{2}\,\partial_k\psi^{(1)} \,D^{ki} {\chi^{(1)}}^{\prime}\,+\,2\,
 \psi^{(1)}\,\partial_k D^{ki} {\chi^{(1)}}^{\prime}\,+\, {\psi^{(1)}}^{\prime} \partial_k D^{ki} \chi^{(1)}
\nonumber \\
&-&\partial_k{\psi^{(1)}}^{\prime} D^{ki} \chi^{(1)}\,+\,\frac{1}{2}
 \,\partial^k \omega^{(1)}\,D^i_{~k} {\chi^{(1)}}^{\prime\prime} \,+\, \Aa \partial^k \omega^{(1)} \,
 D^i_{~k} {\chi^{(1)}}^{\prime} \,-\, 4\Ab \partial_k \omega^{(1)} \,D^{ik} \chi^{(1)} \nonumber  \\
&+& 2\Ac \,\partial_k \omega^{(1)} \,D^{ik} \chi^{(1)} \,
-\,\frac{1}{2}\,\partial_k D^{km}\chi^{(1)} \,D_{~m}^i 
{\chi^{(1)}}^{\prime}
 \,-\,\frac{1}{2} \, \partial_k D_{~m}^i {\chi^{(1)}}^{\prime}\,D^{km} \chi^{(1)}  \nonumber \\
&+&
 \frac{1}{4} \partial^i D_{mk} \chi^{(1)} \,D^{km} {\chi^{(1)}}^{\prime}
 \,+\, \frac{1}{2}\, \partial^i D_{mk} {\chi^{(1)}}^{\prime}\,D^{km} \chi^{(1)}
 \,-\,\frac{1}{2}\,D^{ik}\chi^{(1)} \,\partial_m D^m_{~k} {\chi^{(1)}}^{\prime} \Big) \, , \nonumber \\
\ded {G^0_{~i}}\, &=& \frac{1}{a^2} \Big( - \,\Aa \partial_i \phi^{(2)} \,-\,\partial_i
{\psi^{(2)}}^{\prime} \,-\, \frac{1}{4}
\partial_k\,D^k_{~i} {\chi^{(2)}}^{\prime} \,-\, \frac{1}{4} \La {\chi^{(2)}_i}^{\prime}
\,+\, \frac{1}{4} \La \omega^{(2)}_i \,+8 \Aa \phi^{(1)} \partial_i \phi^{(1)} \\
&+&\,4\,\phi^{(1)}\,\partial_i {\psi^{(1)}}^{\prime} 
+ 2 \,{\psi^{(1)}}^{\prime}\partial_i \phi^{(1)} -\, 4\,{\psi^{(1)}}^{\prime}\partial_i
\psi^{(1)} \,-\, 4\,\psi^{(1)}\,\partial_i
 {\psi^{(1)}}^{\prime} \,+\, \partial_i \phi^{(1)}\,\La \omega^{(1)} \, \nonumber \\
&-&\, \partial_i\partial_k \omega^{(1)} 
\,\partial^k \phi^{(1)} 
+8 \Ac \phi^{(1)}\,\partial_i \omega^{(1)} \,-\, 4 \Ab \phi^{(1)}\,\partial_i \omega^{(1)}
 \, -\, 2\,\Aa \,\partial^k \omega^{(1)}\,\partial_i\partial_k \omega^{(1)}
  \,+\,\La \psi^{(1)}\,\partial_i \omega^{(1)} \nonumber \\
&+& \partial^k \omega^{(1)}\,\,\partial_i\partial_k \psi^{(1)}
 \,-\,\partial_k \phi^{(1)}\,D^k_{~i} {\chi^{(1)}}^{\prime}\,+\,
 \frac{1}{2}\,\partial^k \phi^{(1)} \,D_{ik} {\chi^{(1)}}^{\prime} \,-\,
 \psi^{(1)}\,\partial_k D^k_{~i} {\chi^{(1)}}^{\prime} \nonumber \\
&+&\,
 \frac{1}{2}\,\partial_k\psi^{(1)} \,D^k_{~i} {\chi^{(1)}}^{\prime}
 \,-\, {\psi^{(1)}}^{\prime} \partial_k D^k_{~i} \chi^{(1)} \, -\,\partial_k{\psi^{(1)}}^{\prime} D^k_{~i} 
\chi^{(1)}
 \,+\,\partial_i \omega^{(1)}\, \partial_k
 \partial^m\,D^k_{~m} \chi^{(1)} \nonumber \nonumber \\
&-& 2 \,\Ac\, \partial^k \omega^{(1)}\,D_{ik} \chi^{(1)}
 \,+\,\Ab \partial^k \omega^{(1)}\,D_{ik} \chi^{(1)}  \,+\, \partial^k \omega^{(1)}\,\partial_m\partial_i 
\,D^m_{~k} \chi^{(1)} \nonumber \\
&-&\, \frac{1}{2}\,\partial^m \omega^{(1)} \,\partial_k \partial^k
 \,D_{im} \chi^{(1)}
  \,+\, \frac{1}{2}\,\partial_k D^{km}\chi^{(1)} \,D_{im} {\chi^{(1)}}^{\prime}
 \,+\,\frac{1}{2} \, \partial_k D_{im} {\chi^{(1)}}^{\prime}\,D^{km} \chi^{(1)} \nonumber \\
&-&
 \frac{1}{4} \partial_i D_{mk} \chi^{(1)} \,D^{km} {\chi^{(1)}}^{\prime}
 \, -\, \frac{1}{2}\, \partial_i D_{mk} {\chi^{(1)}}^{\prime}\,D^{km} \chi^{(1)} \Big)\, , \nonumber \\
\ded {{G^{d}}^i_j} &=& \frac{1}{a^2} \Big( +\, \frac{1}{2}\La \phi^{(2)} \,+\,\Aa
{\phi^{(2)}}^{\prime} \,+\,2\,\Ac\,\phi^{(2)} \,-\,\Ab \phi^{(2)} \, -\,\frac{1}{2}\La \psi^{(2)}  \,
+\,{\psi^{(2)}}^{\prime\prime} \\
&+&\, 2\,\Aa\,{\psi^{(2)}}^{\prime}\,+\, \Aa \,\La \omega^{(2)} \,+\, \frac{1}{2}
\La {\omega^{(2)}}^{\prime} \,-\,\frac{1}{4}\,\partial_k\partial_i\,D^{ki}\chi^{(2)}
 \,+\,4\,\Ab \left( \phi^{(1)} \right)^2\nonumber \\
&-&8\,\Ac \,\left( \phi^{(1)} \right)^2\,-\,8\,\Aa\,\phi^{(1)}\,{\phi^{(1)}}^{\prime} \,-\,\partial_k
\phi^{(1)}\,\partial^k \phi^{(1)} \,-\,2 \phi^{(1)}\,\La \phi^{(1)} \,-\, 4\,\phi^{(1)}\,
{\psi^{(1)}}^{\prime \prime}
\nonumber \\
&-&2\,{\phi^{(1)}}^{\prime}{\psi^{(1)}}^{\prime} \,-\,8\,\Aa\,\phi^{(1)}\,{\psi^{(1)}}^{\prime}
\,-\,2\,\partial_k \psi^{(1)} \,\partial^k \psi^{(1)} -\, 4\,\psi^{(1)}\,\La \psi^{(1)}
\,+\, \left( {\psi^{(1)}}^{\prime} \right)^2 \,+\, 8\,\Aa \,\psi^{(1)}\,{\psi^{(1)}}^{\prime} \nonumber \\
&+&4\,\psi^{(1)}\,{\psi^{(1)}}^{\prime \prime}\,+\, 2\,\psi^{(1)}\,\La \phi^{(1)}
\,-\,{\phi^{(1)}}^{\prime}\La \omega^{(1)} \,-\,2\,\phi^{(1)}\,\La {\omega^{(1)}}^{\prime}
\,-\,2\,\Aa\,\partial_k \omega^{(1)}\,\partial^k \phi^{(1)} \nonumber \\
&-&4\,\Aa \,\phi^{(1)}\,\La \omega^{(1)} \,+\, \frac{3}{2}\Ac \partial_k
\omega^{(1)}\,\partial^k \omega^{(1)} \,-\, \Ab
\partial_k \omega^{(1)} \,\partial^k \omega^{(1)} \,+\, 2\,\Aa \partial_k \omega^{(1)} \,\partial^k 
{\omega^{(1)}}^{\prime}\nonumber \\
&-&\frac{1}{2}\,\La \omega^{(1)} \,\La \omega^{(1)}
\,+\,\frac{1}{2}\,\partial^m\partial^k \omega^{(1)}\,\partial_m\partial_k \omega^{(1)}
\,+\, 4\,\Aa\,\psi^{(1)}\,\La \omega^{(1)} \,+\, 2\,\psi^{(1)}\,\La {\omega^{(1)}}^{\prime}\nonumber \\
&-& 2\,\partial_k \omega^{(1)}\,\partial^k
{\psi^{(1)}}^{\prime}\,-\,{\psi^{(1)}}^{\prime}\La \omega^{(1)}
 \,-\, \partial_k\partial_m \phi^{(1)}\, D^{km} \chi^{(1)}  \,-\, \partial_k \phi^{(1)}\,\partial_m D^{mk} 
\chi^{(1)}
 \nonumber \\
&-&\partial_k \psi^{(1)} \,\partial_m D^{mk} \chi^{(1)} \,-\, \frac{3}{2}
\,\partial_k \omega^{(1)} \,\partial^i D^k_{~i} {\chi^{(1)}}^{\prime}
 \,-\,\partial_k {\omega^{(1)}}^{\prime}\,\partial_m D^{mk} \chi^{(1)} \nonumber \\
&-&
\partial_k\partial_m {\omega^{(1)}}^{\prime}\,D^{km} \chi^{(1)} \,-\, 2\,\Aa \partial^k
\omega^{(1)} \,\partial_m D^m_{~k} \chi^{(1)} \,-\, 2\,\Aa \partial_m\partial^k \omega^{(1)}\, D^m_{~k}
\chi^{(1)} \nonumber \\
&+&\frac{3}{4}\,\partial_k\partial^l \,D_{ml}\chi^{(1)}\,D^{km}\chi^{(1)}
\,-\,\frac{1}{2}\,\La\,D_{ml}\chi^{(1)}\,D^{ml}\chi^{(1)}
\,+\,\frac{1}{4}\,\partial_m\partial^k\,D_{lk}\chi^{(1)}\,D^{lm}\chi^{(1)} \nonumber \\
&+&\frac{1}{2}\,\partial_k D_{km}\chi^{(1)} \,\partial^l D^{ml}\chi^{(1)}
\,-\,\frac{1}{8}\,\partial^l D_{km}\chi^{(1)} \,\partial_l D^{km}\chi^{(1)}
+\,\frac{1}{2}\,D^{mk}\chi^{(1)}\,D_{mk}{\chi^{(1)}}^{\prime\prime} \nonumber \\
&+& \frac{3}{8}\, D^{mk}{\chi^{(1)}}^{\prime}\,D_{mk}{\chi^{(1)}}^{\prime} \,+\,
\Aa\,D^{mk}\chi^{(1)}\,D_{km}{\chi^{(1)}}^{\prime} \Big) \delta^i_{~j} \, .\nonumber \\
\ded{{G^{nd}}_j^i}&=& \frac{1}{a^2} \Big[ 
-\frac{1}{2} \,\partial^i\partial_j \phi^{(2)}
\,+\,\frac{1}{2}\,\partial^i\partial_j \psi^{(2)} \,-\, \Aa
\partial^i\partial_j \omega^{(2)} \,-\, \frac{1}{2}\,\partial^i\partial_j
{\omega^{(2)}}^{\prime} \,-\, \frac{1}{2} \Aa \left( \partial^i \,\omega^{(2)}_j +
\partial_j\, \omega^{i(2)}\right) \\
 &-&\, \frac{1}{4} \left( \partial^i \,{\omega^{(2)}_j}^{\prime} + \partial_j\,
{\omega^{i(2)}}^{\prime} \right) \, +\, \frac{1}{2} \Aa \left( D^i_{~j}
{\chi^{(2)}}^{\prime} +
\partial^i {\chi^{(2)}_j}^{\prime}
+ \partial_j {\chi^{i(2)}}^{\prime} + {\chi^{i(2)}_{~j}}^{\prime}\right)
\,+\, \frac{1}{2} \partial_k\partial^i\,D^k_{~j} \chi^{(2)} \nonumber \\
&-&\,\frac{1}{4} \La D^i_{~j} \chi^{(2)} \,-\, \frac{1}{4} \La \chi^{i(2)}_{~j}\,+\,
\frac{1}{4}\left( D^i_{~j} {\chi^{(2)}}^{\prime\prime} +
\partial^i {\chi^{(2)}_j}^{\prime\prime}
+ \partial_j {\chi^{i(2)}}^{\prime\prime} + {\chi^{i(2)}_{~j}}^{\prime\prime}\right)
 \,+ \, \partial^i \phi^{(1)}\, \partial_j \phi^{(1)} \nonumber \\
&+&\, 2\,\phi^{(1)}\,\partial^i\partial_j \phi^{(1)} \,-\, 2\,\psi^{(1)}
\partial^i\partial_j \phi^{(1)}\,-\,
\partial_j \phi^{(1)}\,
\partial^i\psi^{(1)} \,-\, \partial^i \phi^{(1)}\,\partial_j \psi^{(1)} \,+\,
3\,\partial^i \psi^{(1)}\,\partial_j \psi^{(1)} \,+\, 4 \,\psi^{(1)}\,\partial^i\partial_j \psi^{(1)} 
\nonumber \\
&+&\,2\,\Aa\,\partial^i \omega^{(1)}\,\partial_j \phi^{(1)} \,+\, 4 \Aa
\phi^{(1)}\,\partial^i\partial_j \omega^{(1)} \,+\, {\phi^{(1)}}^{\prime}\,\partial^i\partial_j \omega^{(1)}
\,+\, 2\,\phi^{(1)} \,\partial^i\partial_j {\omega^{(1)}}^{\prime} \,+\, \La \omega^{(1)}
\,\partial^i\partial_j \omega^{(1)} \nonumber \\
&-& \partial_j
\partial^k \omega^{(1)}\,\partial^i \partial_k \omega^{(1)}
\,-\,2 \, \Aa \partial^i \psi^{(1)} \,\partial_j \omega^{(1)} \,-\, 2\,\Aa
\partial^i \omega^{(1)}\,\partial_j \psi^{(1)} \,-\, \partial^i
{\psi^{(1)}}^{\prime}\,\partial_j \omega^{(1)} \,+\, \partial_j
{\psi^{(1)}}^{\prime}\,\partial^i \omega^{(1)} \nonumber \\
&-&\, \partial^i \psi^{(1)} \,\partial_j {\omega^{(1)}}^{\prime} \,-\, \partial_j
\psi^{(1)} \,\partial^i {\omega^{(1)}}^{\prime} \,-\, 2\,\psi^{(1)}\,\partial^i \partial_j
{\omega^{(1)}}^{\prime} \,+\, {\psi^{(1)}}^{\prime}\,\partial^i\partial_j \omega^{(1)} \,-\,
4\,\Aa\,\psi^{(1)}\,\partial^i\partial_j \omega^{(1)}   \nonumber \\
&-& 2\,\Aa \phi^{(1)}\,D^i_{~j} {\chi^{(1)}}^{\prime}\,-\,\frac{1}{2} {\phi^{(1)}}^{\prime}\,D^i_{~j}
{\chi^{(1)}}^{\prime} \,-\, \phi^{(1)}\,D^i_{~j} {\omega^{(1)}}^{\prime\prime} \,+\, \frac{1}{2}
\partial_k \phi^{(1)} \, \partial^i D^k_{~j} \chi^{(1)} \,+\, \frac{1}{2} \partial_k \phi^{(1)}
\, \partial_j D^{ki} \chi^{(1)} \nonumber \\
&-& \frac{1}{2} \partial_k \phi^{(1)} \,
\partial^k D^i_{~j} \chi^{(1)} \,+\,\partial_j \partial_k \phi^{(1)}\,D^{ki}\chi^{(1)} \,+\, \frac{1}{2}
{\psi^{(1)}}^{\prime}\, D^i_{~j} {\chi^{(1)}}^{\prime} \,+\,{\psi^{(1)}}^{\prime\prime}D^i_{~j} \chi^{(1)}
\,+\, 2\,\Aa {\psi^{(1)}}^{\prime}\,D^i_{~j} \chi^{(1)} \nonumber \\
&+& \frac{1}{2}
\partial_k \psi^{(1)}\,
\partial^i D^k_{~j} \chi^{(1)} \,+\, 2\,\Aa \,\psi^{(1)}\,D^i_{~j} {\chi^{(1)}}^{\prime}
\,+\, \psi^{(1)}\,D^i_{~j} {\omega^{(1)}}^{\prime\prime} \,+\, \frac{1}{2} \partial_k
\psi^{(1)}\,
\partial_j D^{ki} \chi^{(1)} \,-\,\frac{3}{2}\, \partial_k \psi^{(1)} \,\partial^k D^i_{~j}
\chi^{(1)} \nonumber \\
&+&2\, \psi^{(1)}\,\partial_k\partial^i D^k_{~j} \chi^{(1)} \,+\,2\,
\psi^{(1)}\,\partial_k\partial_j D^{ki} \chi^{(1)}\,-\,2\,
\psi^{(1)}\,\partial_k\partial^k D^i_{~j} \chi^{(1)}\,-\La \psi^{(1)} \,D^i_{~j} \chi^{(1)} \,+\,
\partial^i\psi^{(1)}\,\partial_k D^k_{~j} \chi^{(1)} \nonumber  \\
&+&\,\partial_j\psi^{(1)}\,\partial_k D^{ki} \chi^{(1)} \,+\,
\partial_k\partial^i \psi^{(1)}\,D^k_{~j} \chi^{(1)}
\,+\, \frac{1}{2}\,\partial^i \omega^{(1)}\,\partial_k D^k_{~j} {\chi^{(1)}}^{\prime}
\,+\, \frac{1}{2}\,\partial_k\partial^i \omega^{(1)} \, D^k_{~j} {\chi^{(1)}}^{\prime} \nonumber \\
&+& \frac{1}{2}\,\partial_k\partial_j \omega^{(1)} \, D^{ki} {\chi^{(1)}}^{\prime}
\,-\, \frac{1}{2} \,\partial_k\partial^k \omega^{(1)}\, D^i_j {\chi^{(1)}}^{\prime}
\,+\,\frac{1}{2}\, \partial^k \omega^{(1)}\, \partial^i D_{kj} {\chi^{(1)}}^{\prime}
 \,+\,\frac{1}{2}\, \partial^k \omega^{(1)}\, \partial_j D^i_{~k}
{\chi^{(1)}}^{\prime} \nonumber \\
&-&\partial^k \omega^{(1)} \,\partial_k D^i_{~j} {\chi^{(1)}}^{\prime}\,+\, \frac{1}{2}
\,\partial^k {\omega^{(1)}}^{\prime} \,\partial^i D_{kj} \chi^{(1)} \,+\,
\frac{1}{2} \,\partial^k {\omega^{(1)}}^{\prime} \,\partial_j D_{~k}^i \chi^{(1)}
 \,-\, \frac{1}{2} \,\partial^k {\omega^{(1)}}^{\prime} \,\partial_k D^i_{~j} \chi^{(1)} \nonumber \\
&+& \,\partial_k \partial_j {\omega^{(1)}}^{\prime}\,D^{ik}\chi^{(1)} \,+\, \Aa
\partial^k \omega^{(1)} \,\partial^i
D_{kj} \chi^{(1)} \,+\, \Aa \partial^k \omega^{(1)} \,\partial_j D_{~k}^i \chi^{(1)} \,-\, \Aa
\partial^k \omega^{(1)} \,\partial_k D^i_{~j} \chi^{(1)} \nonumber \\
 &+& 2\,\Aa\,\partial_k \partial_j
\omega^{(1)}\,D^{ik}\chi^{(1)} \,-\,\frac{1}{2}\, D^{ki} {\chi^{(1)}}^{\prime} \, D_{kj}
{\chi^{(1)}}^{\prime} \,-\, \frac{1}{2} \partial^i D_{mj} \chi^{(1)} \,
\partial_k D^{km} \chi^{(1)} \nonumber \\
 &-& \frac{1}{2} \partial_j D_{~m}^i \chi^{(1)} \,
\partial_k D^{km} \chi^{(1)}  \, +\,
\frac{1}{2} \partial_m D^i_{~j} \chi^{(1)} \, \partial_k D^{km} \chi^{(1)}
\,-\,\frac{1}{2} \partial_k\partial^i D_{mj} \chi^{(1)} \, D^{km} \chi^{(1)}  \nonumber \\
&-& \frac{1}{2} \partial_k\partial_j D^i_{~m} \chi^{(1)} \, D^{km} \chi^{(1)} \,+\,
\frac{1}{2} \partial_k\partial_m D^i_{~j} \chi^{(1)} \, D^{km} \chi^{(1)}
\,+\, \frac{1}{2} D^{km} \chi^{(1)} \,\partial^i\partial_j D_{km} \chi^{(1)} \nonumber\\
&+&\, \frac{1}{4} \partial^i D^{mk} \chi^{(1)} \,\partial_j D_{mk} \chi^{(1)}
\,-\,\partial_k\partial^m \,D^k_{~m} \chi^{(1)}\,D^i_{~j} \chi^{(1)} \,-\, \Aa
\,D_{kj}{\chi^{(1)}}^{\prime}D^{ik}\chi^{(1)} \nonumber   \\
&-&\frac{1}{2}\,D_{kj}{\omega^{(1)}}^{\prime\prime}D^{ki}\chi^{(1)} \,-\, \partial_m
\partial_k \,D^m_{~j} \chi^{(1)}\,D^{ki}\chi^{(1)} \,+\,
\frac{1}{2}\,\partial_m\partial^m \,D_{kj}\chi^{(1)}\,D^{ki}\chi^{(1)} \Big]\, , \nonumber 
\eea
where $\ded{{G^{d}}^i_{~j}}$ stands for the diagonal part of $\ded{{G}_j^i}$, 
which is proportional to $\delta^i_{~j}$, and  $\ded{{G^{nd}}^i_{~j}}$ is the non-diagonal contribution.
\section{Perturbing the Klein-Gordon equation}
\label{B}
\def\theequation{B.\arabic{equation}}
In the homogeneous background the Klein-Gordon equation for the scalar field $\varphi$ is 
\be
\label{KG0}
{\varphi_0}^{\prime\prime}+2 \frac{a^{\prime}}{a}{\varphi_0}^{\prime}=-\frac{\partial V}{\partial \varphi} a^2
\ee
The perturbed Klein-Gordon equation at first-order is 
\bea
\label{KG1}
{\deu\varphi}^{\prime\prime}+2\,\Aa {\deu \varphi}^{\prime}
- \La \deu \varphi- {\phi^{(1)}}^{\prime}{\varphi_0}^{\prime}-
 3\,{\psi^{(1)}}^{\prime}{\varphi_0}^{\prime}-\La {\omega^{(1)}} \,{\varphi_0}^{\prime} \,=
- \deu \varphi \,\frac{\partial^2 V}{\partial \varphi^2}\,a^2-
2\,\phi^{(1)}\,\frac{\partial V}{\partial \varphi}.
\eea
At second order we get
\bea
\label{KG2}
&-& \frac{1}{2}{\ded{\varphi}}^{\prime\prime} \,-\,
\,\Aa {\ded{\varphi}}^{\prime} \,+\,\frac{1}{2} \La \ded{\varphi} \,+\,
{\phi^{(2)}}\,{\varphi_0}^{\prime\prime} \,+\, 2\,\Aa {\phi^{(2)}}\,{\varphi_0}^{\prime} \,+\,
\frac{1}{2}\,{\phi^{(2)}}^{\prime}{\varphi_0}^{\prime}\\ 
&+& \frac{3}{2}\,{\psi^{(2)}}^{\prime}\,{\varphi_0}^{\prime} \,+\,
\frac{1}{2}\,\La {\omega^{(2)}} \,{\varphi_0}^{\prime} \,-\,
4\,\left( \phi^{(1)} \right)^2\,{\varphi_0}^{\prime\prime} \,-\, 8\,\Aa \left( \phi^{(1)} \right)^2\,
{\varphi_0}^{\prime}
\,-\, 4\,\phi^{(1)}\,{\phi^{(1)}}^{\prime}{\varphi_0}^{\prime} \nonumber  \nonumber \\
&+& 2\,{\phi^{(1)}}\,{\deu{\varphi}}^{\prime\prime} \, +\,
{\phi^{(1)}}^{\prime}{\deu{\varphi}}^{\prime} \,+\, 4\,\Aa
{\phi^{(1)}}\,{\deu{\varphi}}^{\prime} \,+\,
\partial^k {\phi^{(1)}}\,\partial_k {\deu{\varphi}} \,-\, 6
\,{\phi^{(1)}}\,{\psi^{(1)}}^{\prime}{\varphi_0}^{\prime} \nonumber \nonumber \\
&+& 6\,{\psi^{(1)}}\,{\psi^{(1)}}^{\prime}{\varphi_0}^{\prime} \, +\,
3\,{\psi^{(1)}}^{\prime}{\deu{\varphi}}^{\prime}\,-\, \partial^k
{\psi^{(1)}}\,\partial_k {\deu{\varphi}} \,+\, 2\,{\psi^{(1)}}\,\La \deu{\varphi} \nonumber  \nonumber \\
&-& 2\,{\phi^{(1)}}\,\La {\omega^{(1)}} \,{\varphi_0}^{\prime} \,-\, \partial^k {\omega^{(1)}}\,\partial_k
{\phi^{(1)}} \,{\varphi_0}^{\prime} \,-\,
\partial^k {\omega^{(1)}} \,\partial_k {\psi^{(1)}} \,{\varphi_0}^{\prime}
\,+\, 2\,{\psi^{(1)}}\,\La {\omega^{(1)}} \,{\varphi_0}^{\prime}\nonumber \\
&+& \partial^k {\omega^{(1)}}\,\partial_k {\omega^{(1)}} \,{\varphi_0}^{\prime\prime} \,+\,
2\,\Aa \partial^k {\omega^{(1)}} \,\partial_k {\omega^{(1)}}\,{\varphi_0}^{\prime} \,+\, \partial_k
{\omega^{(1)}} \,\partial^k {\omega^{(1)}}^{\prime}{\varphi_0}^{\prime} \,+\, 2\,\partial^k
{\omega^{(1)}}\,\partial_k \,{\deu{\varphi}}^{\prime}\nonumber \\
&+& 2\,\Aa
\partial^k {\omega^{(1)}}\,\partial_k \,\deu{\varphi} \,+\, \La {\omega^{(1)}}\,{\deu{\varphi}}^{\prime}
\,+\, \partial^k {\omega^{(1)}}^{\prime}\partial_k \deu{\varphi} \,-\, \partial^k
{\omega^{(1)}}\,\partial_i D^i_{~k} {\chi^{(1)}}\,{\varphi_0}^{\prime}\nonumber \\
&-&
\partial_i\partial_k {\omega^{(1)}}\,D^{ik}{\chi^{(1)}}\,{\varphi_0}^{\prime}
\,-\, \partial_i \partial_k \,\deu{\varphi} \,D^{ik}{\chi^{(1)}} \,-\,
\partial_k\, \deu{\varphi}\, \partial^i D^k_{~i} {\chi^{(1)}}  \nonumber \\
&+&
\frac{1}{2}\,D^{ik}{\chi^{(1)}}\,D_{ki}{\chi^{(1)}}^{\prime}\,{\varphi_0}^{\prime} = 
\frac{1}{2} \frac{\partial^2 V}{\partial \varphi^2}\,\ded{\varphi}\, a^2
\,+\, \frac{1}{2}\,\frac{\partial^3 V}{\partial \varphi^3} (\deu{\varphi})^2 \,a^2 \nonumber \, .
\eea
To obtain the Klein-Gordon equation in the longitudinal gauge of Eq.~(\ref{metric3}) one can simply set
$\chi^{(1)}=\chi^{(2)}=0$, and $\phi^{(1)}=\psi^{(1)}$.
Thus at first-order we find
\bea
\label{KG1L}
{\deu\varphi}^{\prime\prime}+2\,\Aa {\deu \varphi}^{\prime}
- \La \deu \varphi- 4 {\phi^{(1)}}^{\prime}{\varphi_0}^{\prime}=
- \deu \varphi \,\frac{\partial^2 V}{\partial \varphi^2}\,a^2-
2\,{\phi^{(1)}}\,\frac{\partial V}{\partial \varphi}\, ,
\eea
while at second order the equation is
\bea
\label{KG2L}
&+& \frac{1}{2}{\ded \varphi}^{\prime\prime} \,+\,\Aa {\ded
\varphi}^{\prime} \,-\frac{1}{2} \La \ded\varphi \,-\, {\phi^{(2)}}\,{\varphi_0}^{\prime\prime}
\,-\, 2\,\Aa {\phi^{(2)}}\,{\varphi_0}^{\prime} \,-\,
\frac{1}{2}\,{\phi^{(2)}}^{\prime}
{\varphi_0}^{\prime}
\\ &-&\, \frac{3}{2}\,{\psi^{(2)}}^{\prime}\,{\varphi_0}^{\prime}
\,-\, 4\,{\phi^{(1)}}\,{\phi^{(1)}}^{\prime}{\varphi_0}^{\prime}
\,-\,4\, {\phi^{(1)}}^{\prime}{\deu \varphi}^{\prime} \,-\, 4\,{\phi^{(1)}}\,\La \deu \varphi \,= \nonumber \\
&-2& {\phi^{(1)}}\,\deu{\varphi}\frac{\partial^2 V}{\partial \varphi^2}\,a^2\,-\,
\frac{1}{2}\ded \varphi\, \frac{\partial^2 V}{\partial \varphi^2}\,a^2
\,-\, \frac{1}{2}\,(\deu \varphi)^2\,\frac{\partial^3 V}{\partial \varphi^3}  \,a^2 \, ,\nonumber
\eea
where we have used the background  equation (\ref{KG0}) and the first-order perturbed equation 
(\ref{KG1L}) to simplify some terms.


\end{document}